**Altimetric measurements of rotating turbulence: cyclone-anticyclone asymmetry, inertial and Kelvin waves and spectral characteristics**

Y. D. Afanasyev[a] and J. D. C. Craig[a, b]

*[a]Memorial University of Newfoundland, St. John's, Canada*

*[b]Science Branch, Dept. of Fisheries and Oceans, NL Canada*

[a]Corresponding author address:

Yakov Afanasyev

Memorial University of Newfoundland, St. John's, NL, Canada

E-mail: afanai@mun.ca

http://www.physics.mun.ca/~yakov

Abstract

Results from a new series of experiments on turbulent flows in a rotating circular container are presented. Electromagnetic forcing is applied to induce flow in a layer of fluid of constant depth. Continuously forced as well as decaying flows are investigated. Optical altimetry is used to measure the gradient of the surface elevation field and to obtain the velocity and vorticity fields with high temporal and spatial resolution. Spectral analysis of the flows demonstrates the formation of dual cascade with energy and enstrophy intervals although the corresponding spectral fluxes of energy and enstrophy are not uniform in these intervals. The energy interval is characterized by the slope of approximately -5/3 in terms of wavenumber and is limited in extent by the finite radius of deformation effect. In the enstrophy range, the slope is steeper than -3 due to the presence of long-lived coherent vortices. The spatial patterns of spectral fluxes in the flow indicate that inverse energy transfer and direct enstrophy transfer occur mainly in elongated vorticity patches. Cyclone/anticyclone asymmetry in favor of anticyclones is observed in our flows. Dominance of anticyclones is most clear during the decay phase of turbulence. The anticyclones remain circular while cyclonic vorticity is stretched into elongated patches. Measurements show that skewness of vorticity distribution increases with increasing Froude number of the flow. Inertial and Kelvin waves are detected in our experiments; their characteristics are analysed against the appropriate dispersion relations.

## I. INTRODUCTION

Turbulent flows in the atmospheres of planets and Earth's oceans are affected by the rotation of the planet to a significant degree. If the background rotation is strong enough compared with the rotation rate of vortices in the turbulent flow, the bulk of the flow tends to become approximately two-dimensional, independent on the coordinate along the axis of rotation, except perhaps in the Ekman boundary layer at the bottom. Despite this significant simplification, rotating turbulence is quite complicated and includes different phenomena that are not completely understood. Previous experiments where the flows in the rotating frame were generated by oscillating grid[1-5], source-sink forcing[6, 7] or electromagnetic forcing[8] addressed, among others, several issues including in particular the process of two-dimensionalization, the asymmetry between cyclonic and anticyclonic vortices and the generation of inertial waves in the flows.

Cyclone-anticyclone asymmetry in nature and the laboratory has been a subject of considerable interest for at least two decades and is yet to be completely understood. The examples of the preference towards anticyclones in nature are well known and include vortices in the atmospheres of gas giants and large scale eddies in the Earth's oceans. The anticyclonic dominance was also observed in numerical simulations of shallow water equations[9-12]. Surprisingly, laboratory experiments on rotating turbulence[3, 4, 13-15] showed a preference of cyclonic rather than anticyclonic vortices. In what follows we will study rotating single-layer fluid without involving stratification or beta-effect to show the asymmetry towards anticyclones.

Yet another phenomenon which contributes to the complexity of rotating turbulence is inertial waves. These occur in rotating systems where Coriolis effect acts as a restoring force. On one hand, inertial waves provide a mechanism of adjustment to two-dimensionality. On the other they are themselves inherently three-dimensional. They are expected to be emitted spontaneously by vortices interacting with each other in turbulent flows, thus inertial waves contribute a three-dimensional component to these flows. Observations of inertial waves in rotating turbulence were reported in the experiments by Bewley et al[5] who detected the resonant modes of inertial waves in the grid-generated turbulence. The resonant modes are determined by the shape and the dimensions of the container and have a discrete spectrum of frequencies. In experiments by Kolvin et al[7], turbulence was generated at the bottom of a tall container by the source-sink method. The authors observed upward propagation of energy of the turbulence and related the speed of the energy front to that of linear inertial waves.

In this paper we present an experimental study of rotating turbulence which is motivated by the geophysical applications. Yet, our setup is idealized and is simple enough such that our experiments can be compared with previous numerical and experimental studies by different authors. Our goal is to contribute to a general picture of the rotating turbulence and to uncover some new phenomenology using the high resolution altimetric measuring technique. We perform our experiments in a layer of large aspect ratio, the ratio of the diameter of the tank to the depth of the layer. The forcing scale to depth ratio is of the order of unity. One can think of this setup as an idealized model of a large-scale quasi-two-dimensional turbulence generated by a medium-scale forcing in a thin atmosphere. However, here we intentionally avoid the β-effect which usually results in formation of zonal flows due to the Rossby wave dynamics[16, 17]. The strength of the background rotation in our flows is somewhat stronger or comparable to the strength of nonlinear terms in the equations of motion such that the Rossby number, $Ro = U/(\Omega L)$ is moderate with $Ro \lesssim 1$ . Here $\Omega$ is the background rotation rate and $U$ and $L$ are velocity and length scales of the flow. The forcing in our experiments is not uniform in the vertical direction and the forcing scale is comparable to the depth of the layer such that the existence of three-dimensional and nonhydrostatic motions such as inertial waves is expected. In terms of control parameters the most important difference between our experiments and previous studies is that the radius of deformation defined as $R_d = (gH)^{1/2}/f_0$, where $g$ is the acceleration due to gravity, $H$ is the depth of the fluid and $f_0 = 2\Omega$ is the Coriolis parameter, is not too large compared to the typical size of vortices in the flow such that vortices "feel" the finite radius of deformation effect. It will be shown in particular that this results in the preference for anticyclones in the flow are in contrast to the results of previous experiments where the dominance of cyclones was reported.

The altimetric system we use for measuring the characteristics of the flow has sufficient temporal and spatial resolution to resolve both turbulent motions and waves. This allows us to analyse the spectral characteristics of turbulence and compare them in greater detail with existing theories and other studies. The concept of two-dimensional turbulence proved to be very useful although it is realized only approximately in natural flows (see review papers by Danilov, Gurarie[18] and Boffetta, Ecke[19]). Kolmogorov type theory was formulated by Kraichnan[20] and developed in further publications by several authors. This theory predicts that purely two-dimensional nonrotating turbulence has a dual cascade where energy is transferred from the scale at which the forcing is applied towards larger scales while enstrophy is transferred towards smaller scales. The corresponding slopes of energy spectrum in terms of wavenumber $k$ are -5/3 in the energy interval and -3 (perhaps with logarithmic correction) in the enstrophy interval. In our experiments we attempt to observe the dual cascade and measure the spectral slopes, examine the frequency domain of the flow and the relation of motions at different scales to inertial waves.

In Sec. II of this paper we describe the setup of our apparatus as well as the optical altimetry technique used to measure the gradient of the surface elevation field and to obtain the velocity and vorticity fields. In Sec. III the results of the experiments and their analyses are reported. Concluding remarks are offered in Sec. IV.

## II. LABORATORY APPARATUS AND TECHNIQUE

The experiments were performed in a circular tank 110 cm in diameter installed on the rotating table (Fig. 1). The tank was rotated in an anticlockwise direction with a rate of $\Omega = 2.32$ $s^{-1}$. A plastic paraboloidal surface with diameter $D = 90$ cm was immersed in the tank. The paraboloid was supported by a cylindrical wall and was concentric with the tank. The paraboloid surface thus constitutes a false bottom of the inner container. The underside of the paraboloid was fitted uniformly at $l_f = 4.6$ cm intervals with about 300 square neodymium magnets $l_m = 2.5$ cm wide and with a field strength of order 1 Tesla. The poles of the magnets were oriented such that their polarity alternated between neighbouring magnets. The shape of the paraboloid was close to that assumed by the surface of the water when in solid-body rotation at the rate $\Omega$. Thus the layer of water with the paraboloidal surface at the bottom was of approximately uniform depth. The depth of the layer $H$ was between 3.8 and 10 cm in different experiments (Table 1). In some experiments the water in the tank was heated up to $55^0$ C in order to reduce its kinematic viscosity. The values of viscosity[21] determined from the measured temperature are given in Table 1 for each experiment. The water was made conductive by dissolving a certain amount of NaCl such that the resulting salinity was between 25 and 35 parts per thousand. Graphite electrodes 30 cm wide and 1 cm high were attached to opposite sides of the paraboloid. The electrodes were placed on the outside of the supporting wall of the paraboloid to preclude bubbles from electrolysis distorting the fluid surface inside the area of interest. Small holes were drilled in the wall to allow electric current to flow through the fluid over the paraboloidal surface. The electrodes were fed current from the unfiltered rectified output of an isolation transformer with a secondary voltage of 117 volts AC. The mean DC current was about 15 amperes. The Lorentz force resulting from the combination of the electric current flowing between the electrodes and the vertical component of the magnetic field acts on the fluid in the horizontal direction perpendicular to the direction of the electric current. The electromagnetic method provides an effective forcing of the fluid in a controlled manner (for more details see Refs 22, 23).

The Altimetric Imaging Velocimetry (AIV) system was used to observe perturbations of the surface topography and to measure two components of the gradient $\nabla \eta = (\partial \eta / \partial x, \ \partial \eta / \partial y)$ of the surface elevation $\eta$ in the horizontal plane $(x, y)$. Here we describe briefly the AIV technique referring for more details to Afanasyev et al.[24] The system includes a 5 Mpix video camera capable of recording with a rate of up to 10 fps and a high brightness computer monitor acting as a light source (Fig.1). A color mask resembling a color wheel used by painters is displayed on the monitor. The video camera observes the reflection of the color mask in the surface of water. When the surface is perturbed by the flow the slopes of the surface elevation field change the angles of reflection. As a result different areas of the light source are reflected. This creates multicolor images of the surface which are used then to visualize the flow and to measure the components of $\nabla \eta$. Note that in the post processing of the measured field of $\nabla \eta$ it was filtered over the scale of 5 pixels to remove the small scale noise and outliers.

Surface velocity of the flow can be determined from the measured gradient of surface elevation using quasigeostrophic approximations

$$\mathbf{V} = \frac{g}{f_0}\left(\mathbf{n} \times \nabla \eta\right) - \frac{g}{f_0^2}\frac{\partial}{\partial t}\nabla \eta - \frac{g}{f_0^3}J\left(\eta, \nabla \eta\right)$$

. (1)

Here $\mathbf{V}$ is the horizontal velocity vector, $\mathbf{n}$ is the vertical unit vector, $g$ is the acceleration due to gravity, $f_0 = 2\Omega$ is the Coriolis parameter and $J(A,B) = \frac{\partial A}{\partial x}\frac{\partial B}{\partial y} - \frac{\partial B}{\partial x}\frac{\partial A}{\partial y}$ is the Jacobian operator. Note that in Eq. (1) and in what follows we use the two-dimensional gradient operator $\nabla = (\partial / \partial x, \partial / \partial y)$. Since Eq. (1) contains the time derivative of $\nabla \eta$, two altimetry images of the flow separated by a short time interval are required to calculate the velocity. The first term in the RHS of Eq. 1 can be easily recognized as the geostrophic velocity. The second and third terms are due to unsteady and nonlinear character of the flow. In the language of dynamical meteorology the latter terms are called isallobaric wind and advective wind respectively. The relative importance of these terms is determined by the temporal Rossby number $Ro_T = 1/\Omega T$ and the Rossby number $Ro = U/\Omega L$ respectively. Here $T$ is the time scale of the flow evolution, while $U$ and $L$ are velocity and length scales of the flow. In the experiments we measure a "true" field of the gradient of the surface elevation (in fact the pressure gradient) which can be expressed in the form of the geostrophic velocity $\mathbf{V}_g = (\mathbf{n} \times \nabla \eta) g / f_0$, and calculate the "total" velocity using equation (1). Note that the calculated total velocity is more accurate when the dimensionless parameters $\mathrm{Ro}_T$ and $\mathrm{Ro}$ are small. In our experiments the values of $\mathrm{Ro}_T$ and $\mathrm{Ro}$ can be of O(1) especially in the centers of strong vortices. Thus we don't expect the total velocity field to be very accurate there.

Inertial waves can also be detected by their surface signature but a different conversion method should be used to reconstruct their velocity field[25]. The conversion is somewhat more complicated; in order to obtain the velocity field, the frequency of the wave has to be measured. This requires a relatively long set of observations (time interval of one or more inertial periods). A Fourier transform can then yield the frequency.

It is often important to calculate the horizontal divergence div$\mathbf{V}$. The horizontal divergence is expected to be small, yet it can be very useful serving as an indicator of 3D effects in the flow. The divergence of the geostrophic velocity field is identically zero assuming that the measured surface elevation $\eta$ is self-consistent. The divergence of the velocity given by Eq (1) is then due to the ageostrophic terms only. However, a straightforward calculation of div$\mathbf{V}$ in the form

$$\mathrm{div}\mathbf{V} = -\frac{g}{f_0^2}\frac{\partial}{\partial t}\nabla^2\eta - \frac{g}{f_0^3}\nabla \cdot J\left(\eta, \nabla\eta\right)$$

$$.$$

(2)

requires numerical differentiation. The measured field $\nabla\eta$ contains noise which is amplified by the finite difference procedure. This limits the utility of Eq. (2) in the calculation of div$\mathbf{V}$. An alternative approach to this problem is to consider the shallow water continuity equation in the following form:

$$H\mathrm{div}\mathbf{V} + \partial\eta/\partial t + (\mathbf{V}\cdot\nabla)\eta = 0. \qquad (3)$$

The total velocity $\mathbf{V}$ can be written as a sum of the geostrophic and ageostrophic velocities, $\mathbf{V} = \mathbf{V}_g + \mathbf{V}_a$, where $\mathbf{V}_a$ includes the isallobaric wind and advective wind as in Eq. (1). It is easy to show that the nonlinear term in Eq. (3) is identically zero for the geostrophic velocity $\mathbf{V}_g$. Thus, the horizontal divergence can be found in the following form:

$$\mathrm{div}\mathbf{V} = -(\partial\eta/\partial t + (\mathbf{V}_a\cdot\nabla)\eta)/H. \qquad (4)$$

The above equation allows us to avoid any additional numerical differentiation in calculating div$\mathbf{V}$. In contrast, it requires integration of the term $\partial(\nabla\eta)/\partial t$ in x- and y-directions in order to obtain $\partial\eta/\partial t$. Integration, however, reduces numerical noise rather than amplifying it.

### III. EXPERIMENTAL RESULTS AND ANALYSES

Here we describe the laboratory experiments in a rotating tank with a uniform depth of the water layer. The flow was forced continuously by an electric current and a regular array of magnets for an extended period of time. After that the forcing was switched off and the flow was allowed to decay. Thus we had an opportunity to observe both forced and decaying turbulence. Let us determine first the parameter space of our flows.

### A. Control parameters

A series of experiments was performed where the depth of the layer, the forcing amplitude and the kinematic viscosity of water were varied (Table 1). These dimensional control parameters determine the regime of the flows which can be characterised by the appropriately defined dimensionless parameters. The Rossby number can be defined as the ratio of the vertical component of relative vorticity, $\zeta = \mathbf{n} \cdot (\nabla \times \mathbf{V})$, to the Coriolis parameter, such that Ro $= \zeta/f_0$. The values of Ro were calculated with the root mean square (RMS) vorticity during the period of forcing and are given in Table 1. Note that the RMS quantities were averaged over a large area of the flow (square with the diagonal equal to the diameter of the flow domain). The range of Ro between the experiments is quite limited despite the fact that forcing was varied in a wider range between 25% and 100% of the maximum. The Rossby numbers are less than 1 but not too small such that they can be considered as moderate. Note also that the Rossby number in strong vortices can exceed 1 as will be shown later. Another important dimensionless parameter is the Reynolds number which can be defined based on the Taylor microscale as Re $= \mathbf{V}^2/(\nu\zeta)$ where $\nu$ is the kinematic viscosity of water. The values of Re calculated with RMS velocity and vorticity are given in Table 1. The Reynolds number varies in the range between 11 and 268. Observations show that the flow with the lowest value of Re = 11 (exp. 4 in Table 1) is only weakly turbulent; the vortices mostly remain in a regular array. At larger Re vortices start moving around interacting with each other such that the flows can be considered fully turbulent without any visible indications of a regular array. A dimensionless number similar to the Reynolds number but based on the bottom (Ekman) friction rather than on ordinary viscosity can be introduced. A ratio of nonlinear term in the equation of motion to bottom friction term is $Re_\alpha$ = $U/(\alpha L) = \zeta/\alpha$, where $\alpha$ is the linear drag coefficient which can be estimated as $\alpha = (\Omega \nu)^{1/2}/H$. Also, the Ekman number can be introduced in a usual way as $Ek = 2\nu/f_0 H^2$. Lastly, it is also important to define the Froude number as the ratio of the fluid speed to the speed of surface gravity waves, $Fr = U/(gH)^{1/2}$. As we will show the free surface effect, characterized by Fr, plays a role in the occurrence of asymmetry between the cyclones and anticyclones in our flows. The values of all dimensionless numbers defined above are given in Table 1.

Table I. Parameters of laboratory experiments.

| Exp. # | $H$, cm | $\nu$, $10^{-2}$ $cm^2/s$ | $U_{rms}$, cm/s | Ro | Re | $Re_\alpha$ | Ek, $\times 10^{-5}$ | Fr, $\times 10^{-3}$ |
|---|---|---|---|---|---|---|---|---|
| 1 | 10 | 1 | 0.58 | 0.24 | 33 | 73 | 4.4 | 6.1 |
| 2 | 10 | 1 | 0.49 | 0.24 | 23 | 73 | 4.4 | 5.1 |
| 3 | 10 | 1 | 0.42 | 0.23 | 17 | 69 | 4.4 | 4.2 |
| 4 | 10 | 1 | 0.33 | 0.21 | 11 | 63 | 5.4 | 3.2 |
| 5 | 9 | 1 | 0.6 | 0.23 | 38 | 63 | 5.4 | 6.8 |
| 6 | 9 | 1 | 0.41 | 0.26 | 14 | 71 | 6.9 | 4.4 |
| 7 | 8 | 1 | 0.7 | 0.23 | 48 | 57 | 6.9 | 8.1 |
| 8 | 8 | 1 | 0.5 | 0.21 | 28 | 51 | 4.6 | 5.8 |
| 9 | 9.8 | 1 | 0.84 | 0.20 | 91 | 59 | 3.6 | 9.3 |
| 10 | 9.8 | 0.8 | 1 | 0.22 | 122 | 73 | 3.4 | 10 |
| 11 | 9.8 | 0.72 | 1.2 | 0.24 | 146 | 82 | 9.9 | 11 |

| 12 | 5.5 | 0.66 | 1.9 | 0.42 | 235 | 84 | 9.9 | 24 |
| 13 | 4 | 0.64 | 2.7 | 0.44 | 268 | 66 | 18 | 30 |
| 14 | 3.8 | 0.56 | 2.1 | 0.43 | 265 | 65 | 17 | 28 |

## B. Observations of the flow evolution

Fig. 2 shows a typical evolution of the flow at relatively high value of Re = 268 (exp. 13 in Table 1). Shortly after the forcing is switched on, a regular array of vortices of alternating sign is formed such that there are approximately 10 vortices of the same sign across the tank. Each magnet generates a spatially localized horizontal force on the fluid and thus induces a vortex dipole[26, 27]. The dipoles induced by neighbouring magnets are directed along parallel lines in the directions opposite to each other such that in the space between two magnets there are two vortices of the same sign. These vortices constitute halves of the dipoles induced by each magnet. Closer inspection of relative vorticity field in Fig. 2 a shows indeed that vortices between the magnets initially have dual cores. However these cores rapidly coalesce into a single vortex. This process can be observed in a vorticity video during the short initial period. Intensity of vortices in the array rapidly grows in time such that the initially regular array of vortices evolves into a turbulent flow where vortices are no longer attached to specific locations (Fig. 2 c, d). The evolution of turbulence comprises vortex formation, translation , interaction and decay by shear. When the forcing is switched off, the flow starts to decay. Fig. 2 e, f shows the vorticity, velocity and surface elevation fields at the end of the experiment. Surprisingly, only anticyclones survived as coherent circular vortices. Their cores can be clearly seen as "black holes" in the vorticity map in Fig. 2 e. Cyclonic vorticity mostly exists in elongated patches and filaments rather than in the form of circular vortices. This observation reveals an asymmetric evolution of cyclones and anticyclones. We will quantify and discuss this effect in more detail later.

Before we proceed to a detailed analysis of the turbulent state of the flow, it is important to establish if the flow is indeed in a statistically steady-state during the period of forcing. In order to do that we measured mean kinetic energy $E = \mathbf{V}^2/2$ and enstrophy $\zeta^2$ of the flows. Time series of both quantities (Fig. 3) show the main features of the flow evolution. A short initial period of about 25 s is characterized by the approximately linear growth of both energy and enstrophy. The energy is "pumped" into the system by the external forcing while the dissipation is yet unable to process all of the incoming energy. The energy of the system "overshoots" reaching a peak then falls down again to a certain extent before reaching a quasi-steady level where energy performs fast fluctuations typical for a turbulent flow. Typical time required to reach a developed turbulent state is about 50 s in this particular experiment (exp. 11 in Table 1) and is close to the characteristic Ekman time as will be shown below.

It is important to quantify dissipation in this system. The effect of viscosity is usually interpreted in terms of "ordinary" viscosity in the bulk of the layer and bottom friction. Bottom friction in the rotating fluid is described by the Ekman layer theory and is commonly parameterised by a linear term in the vorticity equation

$$\frac{D\zeta}{Dt} = -\alpha\zeta + \nu\nabla^2\zeta \ , \tag{5}$$

where $\alpha$ is the linear drag coefficient. The energy budget equation is then[28]

$$\frac{dE}{dt} = -2\alpha E - \nu\zeta^2 . \tag{6}$$

Ekman friction results in exponential decay of the total energy of the flow while the contribution from the ordinary viscosity depends on total enstrophy $\zeta^2$. Energy decay can be easily measured when the forcing is switched off and the flow is allowed to relax ($t > 250$ s in Fig. 3). The insert in Fig. 3 shows that the sum $d(\ln E)/dt + \nu\zeta^2/E$ (lower solid line) approaches the value of -0.04 s$^{-1}$. The bottom friction coefficient is one half of this value and is $\alpha = 0.02$ s$^{-1}$. A theoretical estimate of the bottom drag coefficient, $\alpha = (\Omega\nu)^{1/2}/H$, gives $\alpha = 0.03$ s$^{-1}$ which is in agreement with the measured value. Characteristic Ekman time can be introduced here as $T_E = \alpha^{-1} = 50$ s. The ordinary friction term $\nu\zeta^2/E$ is also shown by upper line in the insert in Fig. 3; ordinary friction is relatively small and is responsible for only 10% of the total energy decay rate. This indicates that the bottom Ekman friction dominates in the process of the energy dissipation.

It is important to discuss here the 3D effects in our flows. Forcing is not uniform across the depth of the layer. The electromagnetic force is applied locally to the fluid in a volume where the vertical component of the magnetic field is significant. The magnetic field above a magnet decays rapidly with distance from the magnet (e.g. Ref. 23) such that the electromagnetic force is only significant across the depth approximately equal to the size of the magnet (2.5 cm in our experiments). Thus the forcing is effectively applied in the lower half or even quarter of the layer in various experiments. The vortices therefore are created by forcing at the lower part of the layer and then extend to entire depth of the layer to form columnar vortices. Thus, the vertical non-uniformity of forcing contributes to both vertical vortex stretching and to the emission of inertial waves during the process of adjustment of vortices. Another essentially 3D effect is the boundary layer at the bottom of the tank. Although the thickness, $d$, of the bottom Ekman layer is quite small, $d = (\nu/\Omega)^{1/2} = 0.05$ cm, the Ekman layer affects the interior of the fluid via a mechanism of Ekman pumping. The Ekman pumping results in the vertical velocity, $w$, in the interior such that this velocity is proportional to the vorticity in the interior

$w = d\zeta / 2 = (\nu\Omega)^{1/2} \text{Ro}$ . The pumping velocity $w$ is directed upward in a cyclonic vortex and downward in an anticyclonic vortex. The magnitude of the vertical velocity is approximately 0.06 cm/s at Ro = 0.5 and is small compared to the RMS horizontal velocity in the flow.

### C. Spectral characteristics

Further insight into the dynamics of the turbulent flow can be provided by an analysis of its spectral characteristics. This approach allows us to investigate the distribution and transfer of energy and enstrophy between different scales of the flow and to gain insight of the universal characteristics of the flow. We performed discrete two-dimensional Fourier transform of the velocity field **V** to obtain the power spectra $E(\mathbf{k}) = \frac{1}{2}|\mathbf{V}(\mathbf{k})|^2$ in wavenumber space $\mathbf{k} = (k_x, k_y)$. The two-dimensional spectrum was then averaged over all possible directions of the wavevector **k** to obtain one-dimensional (isotropic) spectrum defined as $E(k) = 2\pi k \langle E(\mathbf{k}) \rangle$ where $k$ is the magnitude of vector **k** and the average is over all $|\mathbf{k}| = k$. Figure 4 shows one-dimensional spectra for two flows (11 and 13 in Table 1) with somewhat different regimes. Experiment 11 is characterised by lower Reynolds number, Re = 146 and deeper water layer, $H$ = 9.8 cm compared with experiment 13 which was performed with higher Re = 268 and in a shallow layer,

$H$ = 3.8 cm. The lower curves in Fig. 4 a, b show the spectra at the very beginning of each

experiment, only a few seconds after the forcing starts, while the upper curves show spectra

when the (forced) flow is in a statistically steady state.  At the beginning of each experiment the

spectrum is characterized by a sharp peak at the forcing wavenumber $k_f$ at 1 cm$^{-1}$. Note that in the

periodic flow pattern that we observe immediately after the forcing starts (Fig. 2 a, b) we can

count approximately 10 wavelengths across the tank. Thus the forcing wavelength

$\lambda_f \approx D/10 \approx 2l_f \approx 9$  cm, where $D$ is the diameter of the tank and $l_f$ is the distance between the

magnets. The forcing wavenumber in x or y-direction is $k_{x, y} = 2\pi/\lambda_f$  and one-dimensional forcing wavenumber is then $k_f = \sqrt{2}k_{x,y} \approx 1$   cm$^{-1}$. It is interesting to note that a second peak at $k$

= $2k_f$ is also observed in the initial spectra. This peak is related to the double-core vortices that are generated initially by the adjacent magnets as can be seen in Fig. 2 a.

Theory of (nonrotating) two-dimensional turbulence[19] predicts the so-called inverse energy cascade where the energy propagates from the forcing scale towards larger scales or,

alternatively, towards smaller wavenumbers such that $E(k) = C\varepsilon^{2/3} k^{-5/3}$, where $C$ is the

dimensionless Kolmogorov constant and $\varepsilon$ is the energy dissipation rate. The spectra in Fig. 4

show that an interval exists in both experiments where the spectral slope is indeed close to predicted -5/3. The extent of the energy interval in both experiments in Fig. 4 is limited and more so in experiment 14. The limiting factor which determines the outer scale of the energy interval is most likely the finite radius of deformation which is defined as $R_d = (gH)^{1/2}/f_0$. Previous numerical simulations by Polvani et al. showed that coherent vortices which grow beyond $R_d$ do not interact very effectively. This restricts the energy cascade to larger scales. Danilov and Gurarie[18] provide theoretical arguments to this effect. They argue that a velocity field due to a localized potential vorticity anomaly is also localized (the Green function decays exponentially on $R_d$ scale) thus one can expect the lower coherence of vortical structures beyond $R_d$. The radius of deformation depends on the depth of the fluid and is smaller in experiment 13. A relevant dimensionless wavenumber can be introduced as $k_{Rd} = 2\pi/R_d$. The values of $k_{Rd}$ indicated by the arrows in Fig. 4 a, b, correlate well with the size of the energy interval in these experiments. Thus the finite radius of deformation effect provides reasonable explanation for a shorter energy interval in experiment 13.

Bottom friction can also influence the energy cascade. The relevant arrest scale can be introduced from dimensional considerations in the form[29, 18] $L_{fr} = \varepsilon^{1/2}\alpha^{-3/2}$. Estimating $\varepsilon$ as $2\alpha E$, we obtain $L_{fr} = (2E)^{1/2}/\alpha$. The friction scale is typically between 75 and 85 cm in our experiments which is close yet below the size of the domain $D$. This fact has important consequences. If the domain is too small such that $D < L_{fr}$, energy pumped into the system by forcing is not effectively removed by friction. The arrest scale for the energy cascade is then simply the size of the domain. This results in accumulation of energy at the basin scale and formation of a large scale vortex. This regime was conjectured by Kraichnan[20] and was called the condensate regime by analogy with the Bose-Einstein condensation. The large scale vortex significantly affects the motions on smaller scales. The properties of turbulence in this regime were investigated by Xia et al.[30, 31] In our experiments the domain is large enough to avoid condensation yet friction is not too limiting. Thus, we have an optimal combination of these two control parameters which allows us to observe small scale turbulence.

Fig. 4 c and d show compensated energy spectra $\varepsilon^{-2/3} k^{5/3} E(k)$ for experiments 11 and 13 respectively which allow us to estimate the Kolmogorov constant in the energy interval. *C* is approximately equal to 7.5 in experiment 11 and 8.5 in experiment 13. The values of the Kolmogorov constant reported in the literature (mostly from numerical simulations of 2D turbulence) range between 5.8 and 7. The values obtained by Sommeria[32] in his experiment with a layer of mercury range between 3 and 7.5 while Paret and Tabeling[33] found a value of 6.5 in their experiment with a thin layer of salt water. The values of *C* obtained in our experiments for rotating flows are close to the previously reported values obtained in non-rotating simulations or experiments.

The interval extending from the forcing wavenumber towards larger wavenumbers (smaller scales) is called the enstrophy interval. The energy in the enstrophy interval cascades forward to larger wavenumbers (smaller scale) albeith at lower rate than in the inverse cascade in the energy interval. Arguments similar to those used in deriving the spectrum in energy interval give $E(k) \sim k^{-3}$ in the enstrophy interval for nonrotating two-dimensional turbulence. Our observations show spectral slopes that are sligtly steeper than the -3 law to the right of the forcing peak (Fig. 4). The steepening beyond -3 law is consistent with the previous observations in numerical simulations of two-dimensional turbulence[34] and can be explained by the presence of strong coherent vortices. In the forced flow, these vortices are typically at the forcing scale and can be clearly seen in Figs. 2 c, d. Coherent vortices are long-lived and hence they can effectively delay/block the energy transfer to smaller scales that results in a steep spectral slope.

2D turbulence theory predicts only the energy spectrum as a function of wavenumber $k$. In experiments, measurements are often performed at a fixed location to give a time series of velocity. Energy spectrum in terms of frequency, $\omega$ can then be obtained. Although the general theoretical relation between the frequency and wavenumber spectra is not available, the Taylor hypothesis can be used to relate the frequency and wavenumber as $\omega = Uk$, where $U$ is a mean flow advecting the "frozen" turbulence past the location where the measurements are performed. In the absence of mean flow Rhines[34] used RMS velocity of eddies to relate the frequency and wavenumber in order to establish the boundary between turbulence and waves in the turbulent flow on the beta-plane. A recent study by Wunsch[35] gave examples of frequency-wavenumber spectra for the mid-latitude ocean and provided a discussion of the relation between the temporal and spatial structure of the oceanic flows. The frequency-wavenumber spectra can be used for a dual purpose of investigating the properties of the turbulent flow and of identifying waves in the flow. Here we use this approach to investigate the structure of our flows (we will leave the discussion of waves until the following section). We performed a triple Fourier transform (in time and in two horizontal spatial dimensions) of the velocity components of the flow. The time transform was performed for the time interval of about 100 inertial periods ($T_i = 2\pi/f_0 = 1.4$ s) with 7 samples per period. Two-dimensional spatial transform was performed over the entire domain of the flow; the result was then averaged over the directions of the wavevector to obtain a one-dimensional dependence on wavevector magnitude $k$. The averaging procedure was the same as that used to obtain the one-dimensional spectra in Fig. 4. As a result of these transforms we obtain the energy spectrum $E(\omega, k)$ in frequency and wavenumber space. The spectra for two experiments (11 and 13) are shown in Fig. 5. The red lines show linear relation $\omega = Uk$ with the RMS velocity for each experiment. The spectra have a form a plume elongated towards larger $\omega$ and $k$. According to the Taylor hypothesis, if $U$ were the mean advective velocity one would expect the spectral plumes to be aligned along the "nondispersive" lines $\omega = Uk$. However, the lines rather give an upper boundary for the spectral plumes while most of the energy of the flow is at lower frequencies. The spectra indicate that the linear relation between frequency and wavenumber with the RMS velocity does reflect some essential physics although more detailed relation between temporal and spatial structure is needed.

The spectra obtained in our experiments resemble those in 2D turbulence such that one might expect the existence of the corresponding cascades of energy and enstrophy that sustain the spectra. More detailed information about the energy and enstrophy cascades in our flows can be provided by investigation of the corresponding fluxes. Here we adopt a spatial filtering technique which has been proved useful in previous studies of 2D turbulence[36-39]. When energy and enstrophy fluxes are calculated in wavenumber domain using Fourier transform the information about the spatial distribution of the fluxes at different scale (or wavenumber) is lost. In contrast, the filtering technique allows one to resolve the spatial distribution of the fluxes as well as calculate total fluxes across the domain for different scales. The energy flux ($\Pi$) and the enstrophy flux ($Z$) can be defined as follows:

$$\Pi^{(l)} = -\left[\left(v_i v_j\right)^{(l)} - v_i^{(l)} v_j^{(l)}\right] \partial v_i^{(l)} / \partial x_j , \qquad (7)$$

$$Z^{(l)} = -\left[\left(\zeta v_j\right)^{(l)} - v_i^{(l)}\zeta^{(l)}\right]\partial\zeta_i^{(l)}/\partial x_i .  \tag{8}$$

Here $v_i$ is the $i$th component of the horizontal velocity and summation is implied over repeated indices. The superscript $(l)$ denotes filtering at scale $l$. The filtering is performed by convolving the correspondent field with the Gaussian kernel $G^{(l)} = 9/(2\pi\, l^2)\exp(-9r^2/2l^2)$. Thus the field is smoothed with the low-pass filter with cutoff $l$. In order to relate the results of filtering to those obtained by Fourier transforms we can introduce a wavenumber corresponding to the scale $l$ as $k = 2\pi/l$. An example of the smoothed vorticity field is shown in Fig. 6 together with the original field. The filtering is performed at wavenumber $k = 2.1$ cm$^{-1}$. Figure 6 c shows the small scale vorticity obtained by subtracting the filtered field $\zeta^{(l)}$ from the original field $\zeta$ which can be interpreted as a high-pass filtering.

The energy flux Eq. (7) is the scalar product of the stress tensor (the quantity in square brackets) which is due to motions at scale less than $l$ and the strain tensor due to scales greater than $l$. Thus $\Pi^{(l)}$ is the energy flux out of large scales to scales smaller than $l$. While positive values of $\Pi^{(l)}$ indicate transfer to small scales (large wavenumbers), its negative values indicate transfer to large scales (small wavenumbers). The enstrophy flux is defined in a similar manner. Figure 7 shows the snapshots of energy and enstrophy fluxes in experiment 13 (Table 1). The averaging was performed at scale $l = 3$ cm for the enstrophy flux and at larger scale $l = 6$ cm for the energy flux which gives the values of $k$ in enstrophy and in energy interval respectively. In plates (a) and (c) the contours of $\Pi^{(l)}$ and Z$^{(l)}$ are shown in the central part of the flow domain and are superposed on the (greyscale) vorticity map. Plates (b) and (d) show a further magnification of the central area. Vortex structures that are typical in our flows include strong vortices which are usually deformed by strain and are of elliptic shape (a typical vortex is indicated by 1 in Fig. 7) or elongated vorticity patches (indicated by 2 in Fig. 7). Vortices exhibit a distinct quadrupolar pattern of both $\Pi^{(l)}$ and $Z^{(l)}$. Quadrupolar structure of fluxes in vortices was observed recently in the numerical simulations by Xiao et al.[39]. The authors discussed the similarity of the flux structure to the structure of the so-called "palinstrophy production" or vorticity gradient stretching explained analytically by Kimura and Herring[40]. However, it is not obvious if the quadrupolar patterns of fluxes in vortices exhibit any definite shift towards positive or negative total flux. Meanwhile one can expect the predominance of positive enstrophy flux (transfer to smaller scales) and of negative energy flux (transfer to larger scales) to sustain the cascades of enstrophy and energy. The important question therefore is whether we can identify areas of the flow where either positive $Z^{(l)}$ or negative $\Pi^{(l)}$ are dominant. It seems that the areas with elongated patches of vorticity show an asymmetry in the sign of the corresponding fluxes. Such an area is identified by 2 for the enstrophy flux in Fig. 7 b for the energy flux in Fig. 7 d. These observations are consistent with the classical mechanism of the enstrophy transfer to small scales via stretching of vorticity gradients by large-scale strain field[41] or with the so-called vortex thinning mechanism[42] which explains the transfer of energy to larger scales. Both of these mechanisms were discussed in Refs. 36 - 38 on the basis of numerical simulations. It was shown in particular that when a vortex is elongated by the large-scale strain field, a long shear layer is created. In our flows it often involves two shear layers of opposite sign which constitute a jet. As a result of vortex stretching its velocity weakens but the stress produced by the stretching acts to strengthen the large scale strain. Thus the energy of stretched small scale vortex is transferred to large scale field.

The overall balance of energy and enstrophy fluxes can be calculated by averaging the respective fluxes over the domain and over time during the stationary phase of the flow evolution. Figure 8 shows the mean energy flux and enstrophy flux as a function of the averaging wavenumber $k$. The energy flux (lower curve) is negative for all $k$ that indicates the spectral energy transfer to larger scales everywhere although the flux becomes small at small and large wavenumbers. The derivative of $\Pi^{(l)}$ with respect to the wavenumber $k$ is equal to the difference of forcing $E_f$ (energy source) and dissipation $E_d$ (energy sink), $\partial \Pi^{(l)} / \partial k = E_f - E_d$. Thus, the positive slope of $\Pi^{(l)}$ in the range between $k = 1 \text{cm}^{-1}$ and $k = 3 \text{cm}^{-1}$ indicates the presence of forcing in this wavenumber interval. The negative slope at smaller $k$ implies the energy sink. The inverse energy cascade is arrested at small $k$ due to the finite radius of deformation effect in combination with Ekman friction, as discussed before. Although kinetic energy cascades to larger scales at all $k$ there is no definite interval where $\Pi^{(l)}$ is constant. Thus, we cannot expect a clear inertial range in this flow. Finally, the spectral enstrophy flux (upper curve in Fig. 8) is positive except a small interval at small $k$. The predominantly positive $Z^{(l)}$ indicates that the enstrophy cascades to smaller scales as one might expect.

### D. Inertial and Kelvin waves

One of the important features that make rotating turbulence different from its nonrotating two-dimensional counterpart is the presence of certain types of waves in the rotating flow. These waves include in particular inertial waves and Kelvin waves. Although waves are usually associated with three-dimensional motion of fluid they can be detected by their surface signature. For further discussion on observations of inertial waves by their surface signature using altimetry system see Ref. 25. It is, however, usually difficult to separate waves and turbulence. If the vortical flow is assumed to be approximately two-dimensional then the horizontal divergence div**V** can reveal any departures from two-dimensionality. Figure 9 shows the field of div**V** calculated using Eq. (4) in the experiment 13. Typical values of divergence are small, $O(10^{-3})$ compared with vorticity in the flow. Figure 9 b shows a magnified area of the flow and allows us to see the distribution of divergence associated with vortices and jets. Vortices (indicated by 1 in Fig. 9 b) typically form a quadrupolar pattern of divergence while a jet (indicated by 2 in Fig. 9 b) forms a dipolar pattern which contains a converging flow at the jet's entrance and divergent flow at its exit. However, it is unclear if these patterns show waves or they are just the result of the three-dimensionality of vortices. A further inspection of Fig. 9 a reveals a different pattern which almost certainly can be identified with a wave. Large scale pattern of alternating darker and lighter areas can be observed along the boundary of the tank. It can be shown that this pattern corresponds to the Kelvin wave propagating along the boundary. To see the wave more clearly we calculated the divergence due to the term $\partial \eta / \partial t$ in the RHS of Eq. (4) only. This filters out the small scale features due to turbulence such that only large scale pattern of fast Kelvin wave remains. The sequence of images in Fig. 10 shows the propagation of the wave. Kelvin waves are non-dispersive and propagate with speed $c = (gH)^{1/2}$ leaning on the boundary to the right (Northern hemisphere, counter-clockwise background rotation). In the experiment 13 where the layer was relatively deep layer the speed was very high, $c_K \approx 96$ cm/s. The speed of the wave measured by following its crests and troughs in the sequence of images is approximately 93 cm/s and is close to the theoretical value. Three wavelengths fit the circumference of the tank in this experiment such that the wavelength is 94 cm and the wavenumber $k = 0.07$ cm$^{-1}$. The dispersion relation for the Kelvin wave in the form $\omega = ck$ is plotted in the diagram in Fig. 5 together with the measured data point.

Inertial waves have frequency below inertial frequency $f_0 = 2\Omega$ and are ageostrophic and nonhydrostatic three-dimensional motions. In the turbulent flow, waves are emitted by vortices which interact with each other and undergo a process of adjustment. It is quite difficult however to identify inertial waves in a turbulent vortical flow. Bewley et al.[5] observed resonant modes of inertial waves in their rotating turbulence experiments. The resonant modes are determined by the form and the dimensions of the container. The discrete frequencies of standing axisymmetric inertial oscillations in a cylindrical container were obtained by Kelvin and are given by[43]

$$\omega_{mn} = f_0 \left[ 1 + \left( \alpha_{1m} H / \pi n R \right)^2 \right]^{-1/2} , \tag{9}$$

where $H$ and $R$ are the height and the radius of the container respectively, $m = 1, 2 \ldots$ and $n = 1, 2\ldots$ are the radial and axial wavenumbers, and $\alpha_{1m}$ is the m-th zero of the Bessel function of the first order. Note that $m$ can be related to the magnitude of the wave vector as $k = \alpha_{1m} / R$.

Inertial waves can also be described in terms of plane waves propagating at a certain angle to the vertical; the angle of propagation then uniquely determines the frequency of the wave. However, in a relatively shallow layer it is convenient to choose an alternative method where one considers horizontally propagating harmonics in the form $\exp(i(\omega t - k_x x - k_y y))$ with vertical structure determined by their z-dependent amplitudes. Here $\omega$ is the frequency and $\mathbf{k} = (k_x, k_y)$ is the horizontal wavevector with magnitude $k$. The dispersion relation for this waveform is given by[44]

$$-\gamma_n \tan \gamma_n H = \omega^2 / g ,\qquad(10)$$

where

$$\gamma_n = \sqrt{\omega^2 \left(k_x^2 + k_y^2\right) / \left(f_0^2 - \omega^2\right)} ,\; n = 0,\, 1,\, 2\ldots \qquad(11)$$

The dispersion relation (10) in the form $\omega(k)$ is shown as white line in Fig. 5 for the lowest vertical modes $n = 0$ and $n = 1$. The mode $n = 0$ has the lowest frequency for given $k$ and has a simple vertical structure with maxima of horizontal velocity at the surface and at the bottom and zero velocity at mid-depth. The discrete frequencies of the standing resonant modes given by (9) are shown in Fig. 5 by crosses and are located along the line given dispersion relation (10). These frequencies are only shown for the lowest vertical mode in Fig. 5.

The fact that different types of waves have distinct dispersion relations can be used to identify them in a turbulent flow. The lines representing the dispersion relations for Kelvin and inertial waves are superposed on frequency-wavenumber spectra of turbulence in Fig. 5. Kelvin waves are perhaps the easiest to identify because their frequencies are much higher (for a given wavenumber) than those of turbulence. Fig. 5 shows that some energy is concentrated at discrete frequencies along the dispersion line of Kelvin waves. Note that energy at frequencies above $f_0$ and at low $k$ can also be an indicator of inertia-gravity waves with dispersion relation $\omega^2 = f_0^2 + gHk^2$. In our experiments the dispersion relation for inertia gravity waves practically coinsides with that for Kelvin waves above $f_0$.

Inertial waves are harder to identify. Assuming that turbulence is represented by the plume located along (and under) the line $\omega = Uk$ we look for features that extend beyond the plume. In the energy interval at relatively low wavenumbers to the left of the forcing wavenumber the flow contains most energy. Fig. 5 shows that in this interval the energy extends toward higher frequencies beyond the turbulent plume. It seems that this energy extension is bounded from above by the dispersion curve of the inertial waves of the lowest vertical mode. There is also some indication that the discrete modes in $k$ (crosses in Fig. 5) are amplified but the resolution in $k$ or $\omega$ is not enough to identify the resonance modes in our experiments.

Note that the wave energy does not have to be necessarily concentrated at the linear dispersion curve. Firstly, the waves are likely to be nonlinear in the flow with moderate Ro. Secondly, a frequency shift to higher or lower values can occur when inertial waves propagate in the background of either positive or negative vorticity. The inertial waves "feel" local rotation rate of the fluid. The rotation rate in cyclones is higher than in the background. High frequency inertial waves can therefore exist in the cyclones but not in the environment. For similar reason waves with frequency below that specified by their regular dispersion relation (based on background rotation rate) can exist in regions with anticyclonic vorticity. Thus the spectral energy of inertial waves in a turbulent flow can be spread over a wide range of frequencies.

We can retain the information about the spatial distribution of the energy of the motions with different frequencies if we abandon the spatial Fourier transform and only perform the transform in time. Figure 11 shows the resulting spatial distributions of energy for three different frequencies. Zero-frequency, approximately steady-state motions are mapped in Fig. 11 a. These motions are large-scale and their pattern coincides with the pattern of coherent vortices in the flow. The second plate (Fig. 11 b) shows the energy at intermediate frequency, $\omega = 0.3f_0$. This picture reveals relatively small-scale banded features. In the diagram in Fig. 5 these features belong in the area where the spectral plume covers the dispersion relation line at relatively high $k$. They are likely to be small-scale inertial waves emitted by interacting vortices and "living" within vortices. Finally, Fig. 11 c shows the distribution of energy at frequency $\omega = 0.8f_0$. At frequency close to the inertial frequency we can expect to find inertial waves of low wavenumber most likely emitted by most energetic large vortices in the energy interval. The energy distribution reveals indeed large scale features which correspond to large vortices that are also visible in zero-frequency map in Fig. 11 a.

### E.  Cyclone-anticyclone asymmetry

An important feature of rotating turbulence at moderate values of the Rossby number is an asymmetry between cyclones and anticyclones. Visual inspection of vorticity distributions during the forcing period reveals some preference towards anticyclones. Anticyclones keep their circular form while cyclones are easily distorted and stretched into elongated vorticity patches. This asymmetry is more obvious in flows with larger Froude number (Fig. 2 c). The asymmetry towards anticyclones becomes overwhelming during the decay phase where mostly anticyclones survive in the form of closed almost circular vortices (Fig. 2 e). A comparison between the vorticity and the geostrophic vorticity fields reveals further interesting details of asymmetry. Here vorticity is calculated from the (total) velocity **V** while geostrophic vorticity is calculated using the geostrophic velocity $\mathbf{V}_g$. In fact geostrophic vorticity shows a pressure Laplacian. Fig. 12 shows these vorticity fields in experiment 11 in both forcing and decay periods. The comparison shows that during the forcing period the cyclonic vorticity is somewhat stronger in the geostrophic field (Fig. 12 a) while anticyclonic vorticity is stronger in the total field (Fig. 12 b). Note that this only concerns the magnitude of vorticity. The general features including the pattern of vortices (whether they are circular or in the form of elongated patches) and the ratio of cyclones to anticyclones, are the same for both fields. During the decay period the differences between the fields are less perceptible than during the forcing period.

We now quantify the asymmetry and offer an explanation of observed differences in vorticity fields. For quantitative characterization of statistics of vorticity distribution skewness and kurtosis, $S = \left\langle \zeta^3 \right\rangle / \left\langle \zeta^2 \right\rangle^{3/2}, \quad K = \left\langle \zeta^4 \right\rangle / \left\langle \zeta^2 \right\rangle^2,$ are commonly used. Skewness is the third moment of vorticity distribution and shows the asymmetry in the distribution while kurtosis, the fourth moment, is a measure of flatness of the distribution. A typical time dependence of $S$ and $K$ are shown in Fig. 13. During the forcing period between $t = 0$ s and 260 s both $S$ and $K$ are statistically steady. Geostrophic skewness (black line) is positive which indicates a shift towards cyclones while to total skewness (black line) is negative (shift towards anticyclones). When the forcing is switched off both lines of skewness go down towards negative values before they go back to zero at the very end of the experiment. Strongly negative skewness is clearly a consequence of formation of coherent anticyclones during the decay of the flow. Note that we also observed some injections of vorticity from the boundary layer at the sidewall of the container (narrow filaments of vorticity of both sign are visible in Fig. 12 c, d). The injections of relatively strong positive and negative vorticity in approximately equal amount can affect the skewness by driving it towards zero. The values of kurtosis calculated for both geostrophic and total vorticity are not very different from one another during the forcing period and are approximately 3.7 and 3.2 respectively. Thus, our vorticity distributions are just a little bit more flat than the Gaussian distribution where $F = 3$.

We measured mean values of skewness (both based on geostrophic vorticity and total vorticity) during the forcing period in all experiments. Fig. 14 shows skewness as a function of Ro (Fig. 14 a) and Fr (Fig. 14 b). The geostrophic skewness, $S_g$, is positive (in favor of cyclonic vorticity) while the total skewness, $S$, is negative (in favor of anticyclonic vorticity) in all experiments. The magnitude of skewness increases with increasing Ro and Fr. Although scatter is significant, a linear fit can be used (solid lines in Fig. 14) to approximate the data that gives

$S = -1.1\text{Ro} + 0.06$, $S_g = 1.6\text{Ro} - 0.2$,

$S = -10.5\text{Fr} - 0.1$, $S_g = 17\text{Fr} + 0.2$.

Note in particular that the slope of -10.5 in the Froude number dependence of skewness is very close to the value of -10.23 given by Polvani et al.[11] It is quite surprising given the differences in the general setup of our experiments and that of the numerical simulations by Polvani et al. Also, the Froude number in our experiments is quite small and is at the lowest end of the Froude number interval in their simulations.

In order to understand the difference between the geostrophic and total vorticity, consider the process of vorticity generation in our flows. Localized force applied by each magnet generates a symmetric vortex dipole[26, 27]. This vorticity field is independent of background rotation if one ignores a vorticity stretching term $(f_0 + \zeta)\mathrm{div}\mathbf{V}$ in the shallow-water-type equation for relative vorticity. Our measurements show that the horizontal divergence is quite small compared with vorticity which supports this assumption. The pressure field which corresponds to a vortex dipole includes low pressure centers inside both vortices. The main balance in the rotating fluid is geostrophic. Therefore, an additional pressure gradient should be created to balance the Coriolis force which is due to the dipolar flow. The additional pressure is symmetric because the dipole is symmetric and corresponds to high pressure in anticyclones and low pressure in cyclones. The total pressure which we measure in our altimetry experiments is the sum of two fields and is shifted towards negative values. This fact explains the shift towards cyclonic vorticity in the geostrophic vorticity field. A simple model can be offered to quantify this effect. Suppose the velocity field in the dipole created by the forcing is $\mathbf{V}$ and corresponding vorticity and stream function are defined as $\zeta = \mathbf{n} \cdot (\nabla \times \mathbf{V})$ and $\mathbf{V} = -\mathbf{n} \times \nabla \psi$ respectively. The pressure in the dipole is then

$$\frac{p_0}{\rho} = -\frac{V^2}{2} - \int \zeta(\psi) d\psi \,. \tag{12}$$

Let us assume for simplicity that vorticity is constant, $\zeta = \zeta_0$, within each vortex in the dipole. The pressure then takes a simple form[43] $p_0 / \rho = -V^2 / 2 - \zeta_0 \psi$. An additional pressure gradient required to compensate the Coriolis force occurring due to the velocity field in the dipole is given by the geostrophic equation, $\nabla p_1 / \rho = f_0 \left( \mathbf{n} \times \mathbf{V} \right)$. Altimetry measures the geostrophic velocity $\mathbf{V}_g = \left( \mathbf{n} \times \nabla p \right) / \rho f_0$ where the total pressure is $p = p_0 + p_1$. This gives the geostrophic velocity in the form

$$\mathbf{V}_g = \mathbf{V}(1 + \zeta_0 / f_0) - \left( \mathbf{n} \times \nabla V^2 / 2 \right) / f_0 . \tag{13}$$

Geostrophic vorticity can then be found as follows

$$\zeta_g = \mathbf{n} \cdot \left( \nabla \times \mathbf{V}_g \right) = \zeta_0 (1 + \zeta_0 / f_0) - \nabla^2 V^2 / 2 = \zeta_0 (1 + \zeta_0 / 2 f_0) . \tag{14}$$

In the above equation we also used the solid-body-type velocity profile $V = r\, \zeta_0 / 2$ where $r$ is the radial distance measured from the center of a vortex. Eq. (14) shows that geostrophic vorticity has positive (cyclonic) shift by Ro/2 with respect to "true" vorticity of the flow. Averaging over the positive and negative vortex of the dipole one can obtain the skewness shift of approximately 3 Ro /2. If Gaussian distribution of vorticity is assumed in the flow domain, one obtains the skewness shift of 9 Ro /2. The measured difference between the skewness calculated using geostrophic vorticity and that calculated using total vorticity is between these values and is approximately 2.7Ro (Fig. 14 a).

Another way to look at the difference between the geostrophic velocity and total velocity is to consider the balance in the form of the gradient wind relation

$$\mathbf{n} \times \mathbf{V} (\kappa V + f_0) = -g \nabla \eta , \tag{15}$$

where $\kappa$ is the local curvature of the streamlines. Curvature is a scalar field given by the magnitude of the vertical component of the curl of the normalized velocity vector, $\kappa = \mathbf{n} \cdot \left[ \nabla \times (\mathbf{V}/V) \right]$. It has the same sign as vorticity but different in magnitude because the definition of curvature only includes the information about the direction of the velocity vector rather than its magnitude. A typical image of the curvature field in the flow is shown in Fig. 15. The curvature gives an insight into the topology of the flow. A magnified image in Fig. 15 b shows clearly the critical points of the velocity field. Centers of vortices are elliptical critical points which can be seen as white (cyclones) and black (anticyclones) dots while hyperbolic critical points where white and black lines intersect can be seen between vortices. The gradient wind relation includes a centripetal acceleration term which is quadratic in velocity. This term approximates nonlinearity in approximately circular vortices being close to a steady state and with negligible radial velocity. Using the definition of the geostrophic velocity in RHS of eq. (15) we find

$$\mathbf{V}_g = \mathbf{V}\left(1 + \kappa V / f_0\right) \tag{16}$$

The above equation shows that the magnitude of $\mathbf{V}$ is always greater than the magnitude of $\mathbf{V}_g$ in anticyclones and vice versa in cyclones. Assuming again the solid-body-type rotation in vortex cores we obtain relation between the geostrophic and total vorticity, $\zeta_g = \zeta_0(1 + \zeta_0 / 2f_0)$ which is identical to eq. (14).

To illustrate the relation between geostrophic and total velocity in vortices we chose two almost circular vortices in the flow shown in in Fig. 15 and measured the azimuthal and radial velocities using a polar coordinate system for each vortex (Fig. 16). Azimuthal velocity (Fig. 16 a, d) in both vortices is fairly axisymmetric. Radial velocity (Fig. 16 b, e) has quadrupolar distribution which is due to the elliptical deformation of vortices. Note that in general radial velocity is not negligible compared to the azimuthal velocity but its average in azimuthal direction is very small at all $r$. Averaging in azimuthal direction gives radial profiles of the geostrophic and total azimuthal velocity (Fig. 16 c, f). The profiles show clearly that $|\mathbf{V}| > |\mathbf{V}_g|$ in anticyclones and $|\mathbf{V}| < |\mathbf{V}_g|$ in cyclones as predicted by eq. (16).

It is clear why the geostrophic vorticity is shifted towards positive values with respect to total vorticity, yet there is no complete understanding of the preference towards anticyclones in the total vorticity field. Among the arguments for the asymmetry offered by different authors (e.g.) the most applicable to our case is the one offered by Polvani et al.[11] This argument applies to shallow water on the f-plane. Defining potential vorticity (PV) for an axisymmetric vortex as

$q = \left[ f_0 + \dfrac{1}{r} \dfrac{\partial}{\partial r} (rV) \right] / (H + \eta)$ and using it to eliminate the gradient of the surface elevation in the

RHS of the gradient wind equation (15) we obtain

$$\frac{\partial}{\partial r}\left( \frac{1}{r}\frac{\partial}{\partial r}(rV) \right) + \frac{q}{g}\left( f_0 + \frac{V}{r} \right)V = H\frac{\partial q}{\partial r}. \tag{17}$$

The above equation allows one to find velocity for given PV. A scale within which the velocity field "feels" the PV (in Green function sense) is given by

$$R_{in} = \left[ \left( f_0 + V/r \right) q / g \right]^{-1/2} = R_d \left[ \left( 1 + V/f_0 r \right)\left( 1 + \zeta/f_0 \right) \right]^{-1/2}. \tag{18}$$

Since PV is smaller in anticyclones their radius of internal coherence is larger than that for cyclones. Polvani et al.[11] then argue that in cyclones where the radius of internal coherence is small the flow becomes parallel rather than circular. This seems to be a reasonable scenario of stretching cyclones into elongated bands while the anticyclones remain circular.

## IV. DISCUSSION AND CONCLUSIONS

The laboratory experiments reported in this article investigate turbulent flow in a circular container of large aspect ratio subject to background rotation. The optical altimetry method provides us with sufficient resolution to observe the formation of dual cascade of energy and enstrophy in a wavenumber space. Energy interval is characterized by the slope of approximately -5/3 and has its outer limit (at small wavenumbers) constrained by the finite radius of deformation effect. In the enstrophy range the spectral slope is steeper than -3 due to the presence of long-lived coherent vortices that is consistent with observations of other authors. The analysis of the spectral fluxes of energy and enstrophy was performed using spatial filtering. The snapshots of the distribution of the fluxes demonstrate typical quadrupolar patterns in strong circular vortices and predominance of either negative energy flux (transfer to larger scales) or positive enstrophy flux (transfer to smaller scales) in the areas where vorticity patches are elongated. The overall balance of energy flux is negative over the wide range of scales which indicates transfer of energy towards small wavenumbers (inverse cascade). However, the energy flux is not constant in the energy interval. The enstrophy flux is mainly positive which confirms the expectation that enstrophy is transferred to larger wavenumbers.

The rotating system supports waves including inertial waves and, since the vertical boundary is present, Kelvin waves. Fourier transform in time and space allowed us to see the distribution of energy in frequency and wavenumber space. This distribution shows the plume-like dispersion of turbulent energy and reveals some features that can be interpreted as those due to wave motion. Appropriate dispersion relations were used to locate waves in frequency and wavenumber space.

We performed an investigation of the asymmetry between cyclonic and anticyclonic vortices and found that in general the balance is shifted in favor of anticyclones. The asymmetry is not very strong during the forcing period because forcing initially creates symmetric distribution of vorticity. However, anticyclones dominate when the forcing is switched off and the flow is allowed to decay. They maintain their circular form while cyclonic vorticity forms elongated patches. The asymmetry becomes stronger with increasing Froude number such that the absolute value of vorticity skewness increases linearly with Fr with slope of approximately 10.5 which is in a good agreement with previous numerical results by Polvani et al.[11] In contrast, the measurements of geostrophic vorticity demonstrate stronger cyclonic vorticity. Geostrophic vorticity is in fact the Laplacian of pressure or the surface elevation. We showed that pressure field is shifted towards negative values because forcing initially creates negative pressure in both positive and negative vortices.

Our observations of asymmetry of vorticity distribution towards anticyclones contrast with the previous results obtained in experiments with rotating turbulence where the preference towards cyclones was found[3, 4, 13-15]. Thus, our experiments are in distinctly different regime in terms of the dimensionless control parameters. The main reason for this difference is, we believe, the finite radius of deformation effect in our experiments. Moisy et al.[15] argue that the main reason that cyclones are favored over anticyclones in their experiments is the preferential vortex stretching. A vertical strain $\partial w / \partial z$ acts on absolute vertical vorticity $(\zeta + f_0)\mathbf{n}$ such that the stretching is more significant for cyclones (higher absolute vorticity) than for anticyclones (lower absolute vorticity). The stretched vortices are amplified. Direct measurements of the horizontal divergence in our experiments showed that it is very small compared to vorticity. The stretching factor $\partial w / \partial z$ which is related to horizontal divergence as $\partial w / \partial z = -\text{div} \mathbf{V}$ is therefore very small. On the other hand, the radius of deformation in the experiment by Moisy et al.[15] is very large (approximately 30 m) because the depth of the fluid is large and the background rotation rate is small. Thus the finite radius of deformation effect is hardly significant in their experiments to shift the balance towards anticyclones.

Centrifugal instability is commonly thought to be a strong limiting factor on the strength of anticyclonic vortices. The modified Rayleigh criterion[45] predicts that anticyclonic vortices may be subject to inertial instability when their Rossby number exceeds unity. Inertial instability takes the form of toroidal motions of the mushroom-shaped vertical cross section wrapped around the vortex core[46, 47]. The centrifugal instability is very fast; unstable vortices "explode" expanding in the horizontal direction and become weaker such that their Rossby number drops to values below unity[48]. Sreenivasan and Davidson[49] recently showed that the columnar vortices are more likely to form from a cyclonic vertically localized vertical patch than from the anticyclonic due to the mechanism related to centrifugal effects which cause the "radial bursting" of anticyclonic patches at Ro ~ 0.5. In our experiments, however, we routinely observed strong anticyclones with Ro ~ 1 and even somewhat higher. We don't have an immediate explanation of this effect and it certainly deserves further investigation.

Performing the discrete Fourier transform of velocity fields in time and in space we attempted to reveal inertial waves in our turbulent flows. While it is clear that inertial waves are emitted by vortices and should be in abundance in a turbulent flow, it is not trivial to separate them from the vortical component of the flow. The results of our analyses show that inertial waves of relatively low frequency are present at small scale while the waves of frequency close to the inertial frequency exist at the large scale of the most energetic vortices. The energy of inertial waves is however quite small compared with the energy of very low frequency vortical component of the turbulent flow.   Due to the presence of the vertical wall in the tank Kelvin waves occur in our flows. They are very fast waves propagating with gravity wave speed in counter clockwise direction along the wall. They are easily detected in the maps of the horizontal divergence field but their energy is also quite small compared to the turbulent kinetic energy.

**ACKNOWLEDGMENTS**

YDA gratefully acknowledges the support of the Natural Sciences and Engineering Research Council of Canada.

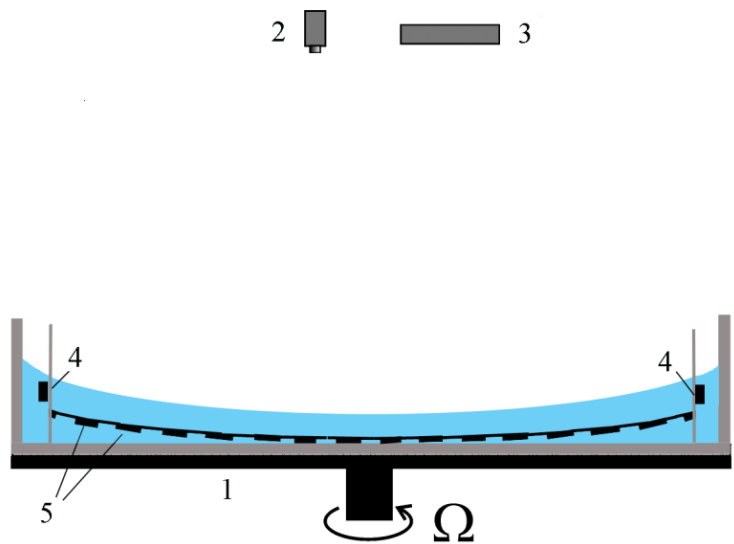

Fig. 1. (Color online) Sketch of the experimental setup: rotating tank and the inner container with the paraboloid false bottom filled with water (1), video camera (2), high brightness TFT panel displaying the color mask (3), electrodes (4) and permanent magnets (5).

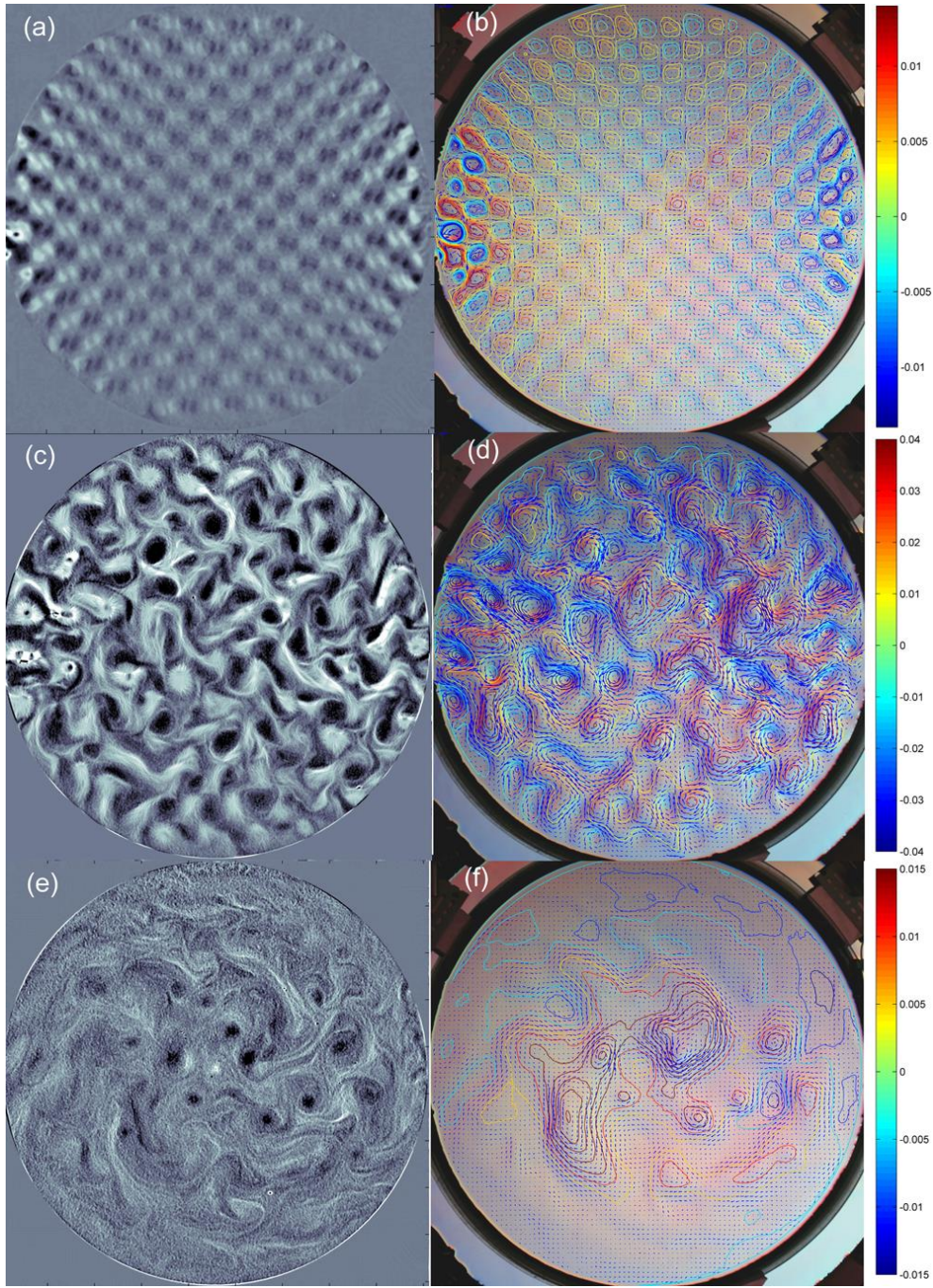

Fig. 2 (Color online) Evolution of the flow in experiment 13 (Table 1): initial period shortly after the forcing starts, $t = 2$ s (a, b), statistically steady forced turbulence, $t = 287$ s (c, d) and decaying turbulence, $t = 28$ s after the forcing is stopped (e, f). Plates a, c and e show the dimensionless vorticity $\zeta/f_0$ in the range from $-1$ (black, anticyclonic) to 1 (white, cyclonic).

Plates b, d and f show velocity vectors superposed on the altimetric images of the flow and contours of surface elevation, $\eta$. The scale shows $\eta$ in cm.

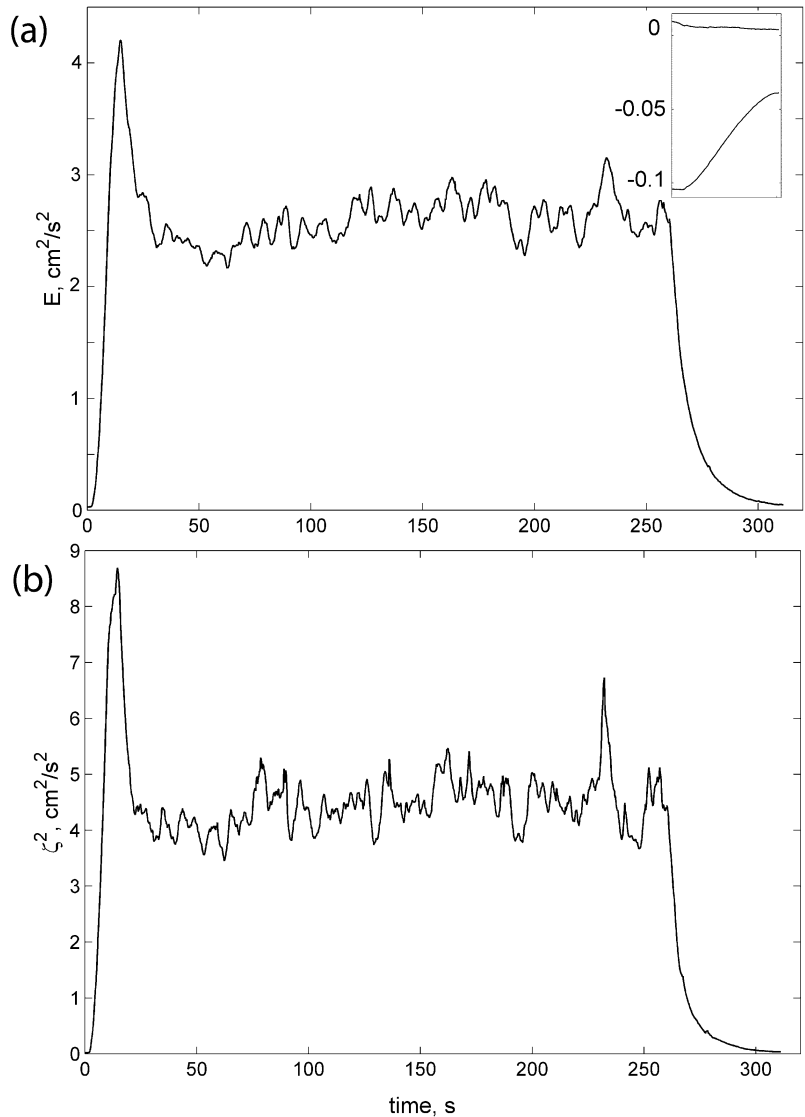

Fig. 3. Kinetic energy $E$ and enstrophy $\zeta^2$ *versus* time in experiment 13 (Table 1). Insert in plate (a) shows $\nu\zeta^2 / E$ (upper solid line) and $d(\ln E) / dt + \nu\zeta^2 / E$ (lower solid line) in s$^{-1}$ after the forcing is switched off.

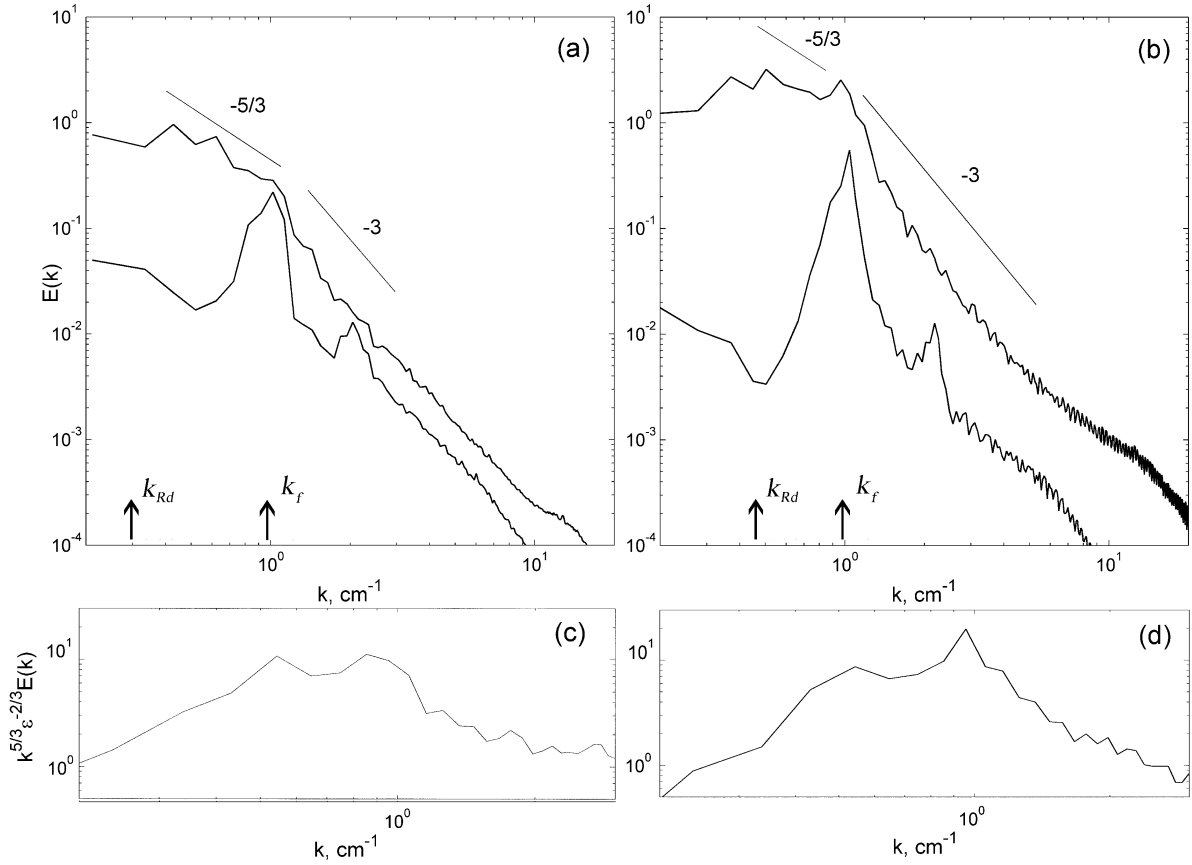

Fig.4. One-dimensional energy spectra of rotating turbulence in two experiments with relatively deep and shallow fluid layer: (a) experiment 10 (Table 1), $t = 3$ s and 390 s; (b) experiment 13 (Table 1), $t = 3$ s and 258 s. Plates c and d show compensated spectra for the experiments 10 and 13 respectively.

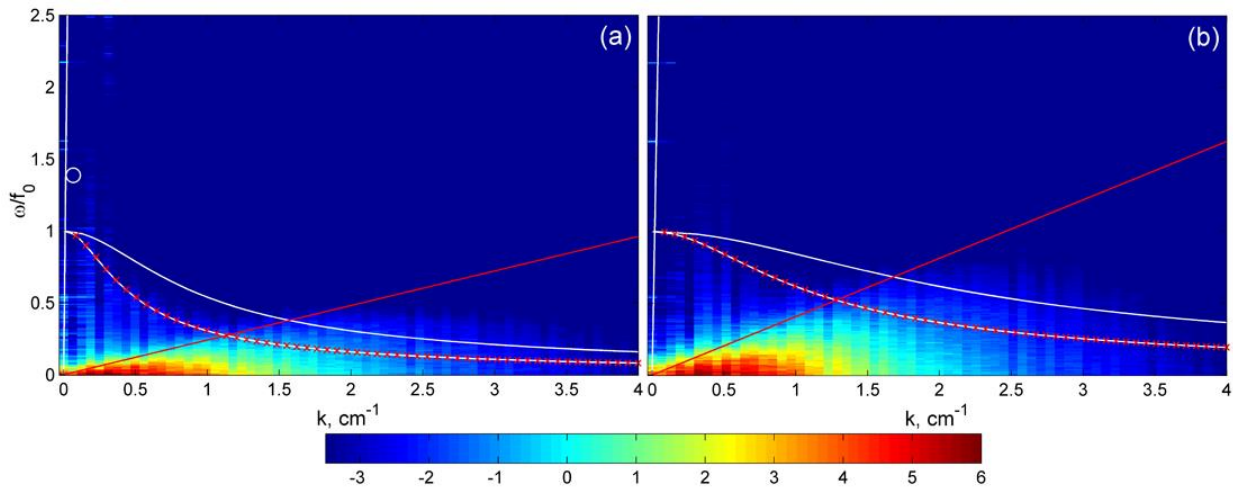

Fig. 5. (Color online) Energy spectra of rotating turbulence shown in frequency and wavenumber space in experiments 11 and 13 (Table 1). Color shows kinetic energy in logarithmic scale. The frequency is normalized by the Coriolis parameter. The solid white lines show the dispersion relations for the inertial waves for the vertical modes *n* = 0 (lower curve) and *n* = 1 (upper curve) as well as the dispersion relation for the Kelvin wave (almost vertical line). Red crosses show discrete frequencies of standing inertial oscillations of lowest vertical mode in the cylindrical container. Red lines show linear relation between frequency and wavenumber with a slope given by the RMS velocity for each experiment (Table 1).

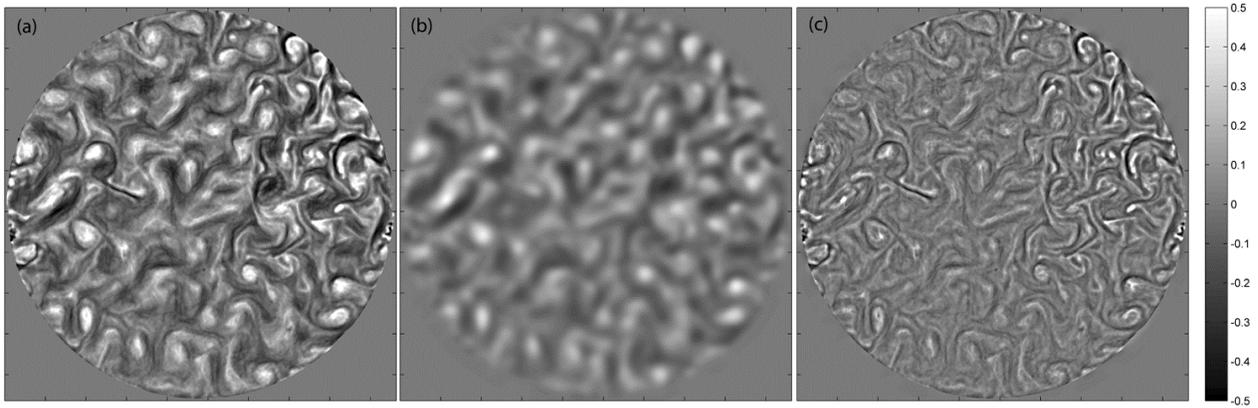

Fig. 6. Vorticity field in experiment 10 (Table 1) at $t = 400$ s: (a) original vorticity $\zeta$, (b) vorticity $\zeta^{(l)}$ filtered by low-pass Gaussian filter with cutoff $l = 3$ cm (wavenumber $k = 2.1$ cm$^{-1}$), (c) small scale vorticity $\zeta - \zeta^{(l)}$. Greyscale shows vorticity normalized by the Coriolis parameter $f_0$.

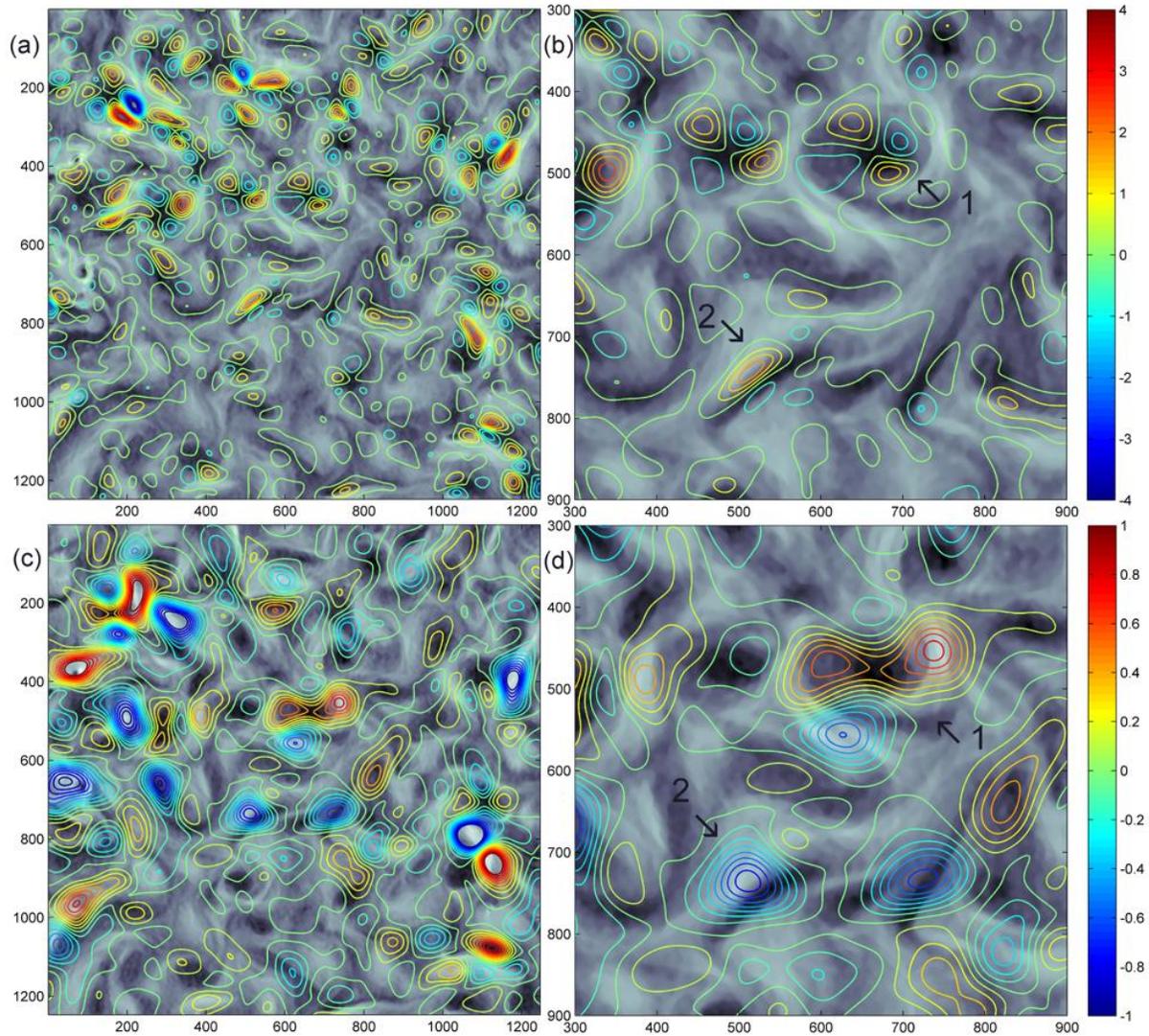

Fig. 7. (Color online) Enstrophy flux $Z^{(l)}$ (a, b) and energy flux $\Pi^{(l)}$ (c, d) fields in the experiment 13 (Table 1) during the forcing period at $t = 258$ s. Spatial distribution of $Z^{(l)}$ in the central area of the flow domain is shown in (a) and further magnified in (b). $Z^{(l)}$ is calculated with $k = 2.1$ cm$^{-1}$ ($l = 3$ cm), scale shows $Z^{(l)}$ in s$^{-3}$. Isocontours of $Z^{(l)}$ are superimposed on the vorticity map where vorticity is normalized by the Coriolis parameter $f_0$ and varies from -1.5 (black) to 1.5 (white). Distributions of $\Pi^{(l)}$ are shown in (c) and (d). $\Pi^{(l)}$ is calculated with $k = 1.05$ cm$^{-1}$ ($l = 5$ cm), scale shows $\Pi^{(l)}$ in cm$^2$ s$^{-3}$. Almost circular vortex with quadrupolar distribution of flux are indicated by 1 while 2 shows an area of elongated vorticity with predominantly positive enstrophy and negative energy fluxes.

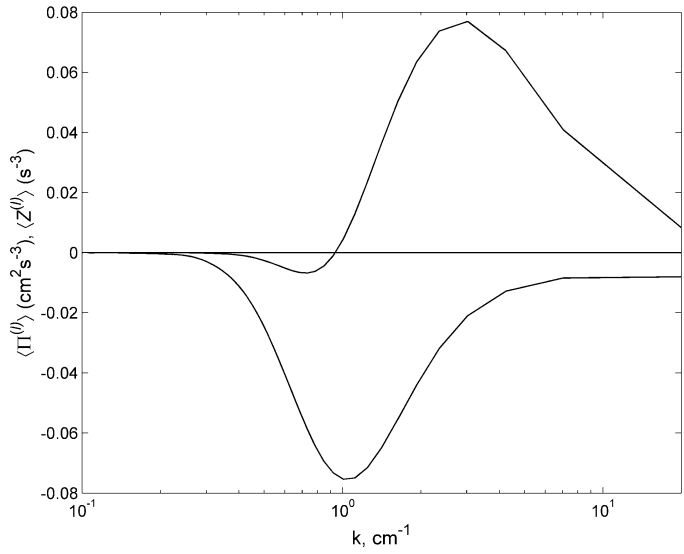

Fig. 8. Mean energy flux (lower curve) and enstrophy flux (upper curve) as a function of wave number based on the averaging scale *l*.

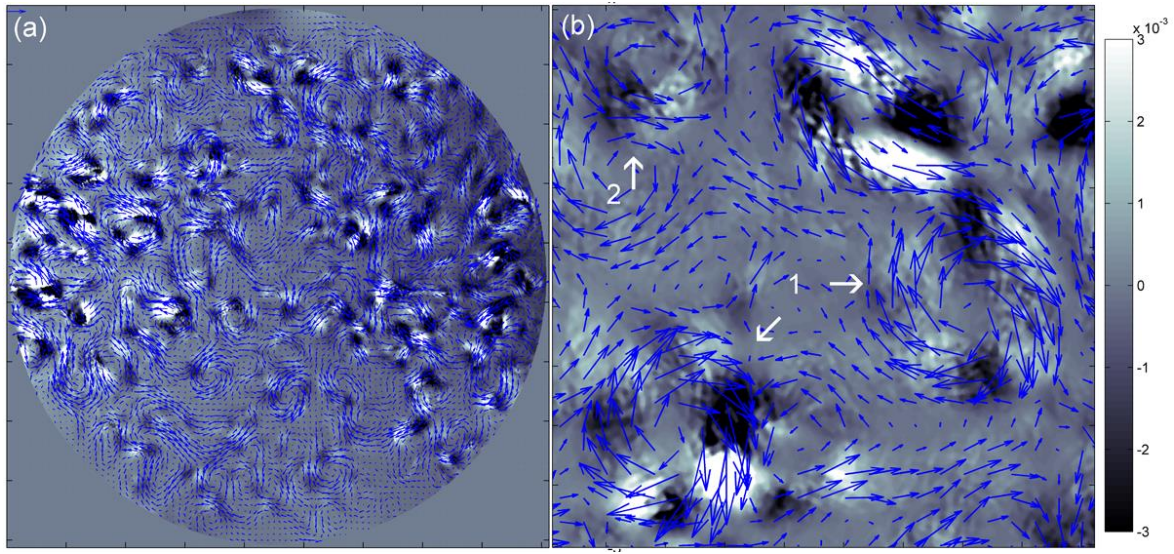

Fig. 9. (Color online) Horizontal divergence and the velocity field in the experiment 13: (a) the entire flow domain, (b) magnified area of the domain showing vortices (1) and a jet (2) between the vortices. Greyscale shows divergence in s$^{-1}$.

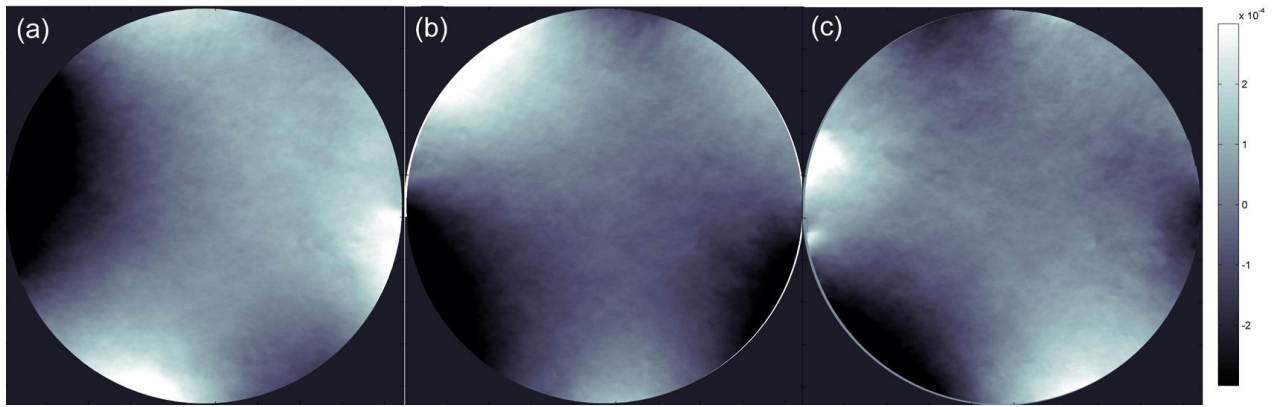

Fig. 10. Sequence of images showing the alternating pattern of divergence associated with the Kelvin wave propagating counter clockwise in the experiment 13. The time interval between the successive images is 0.2 s. Greyscale shows divergence in s$^{-1}$.

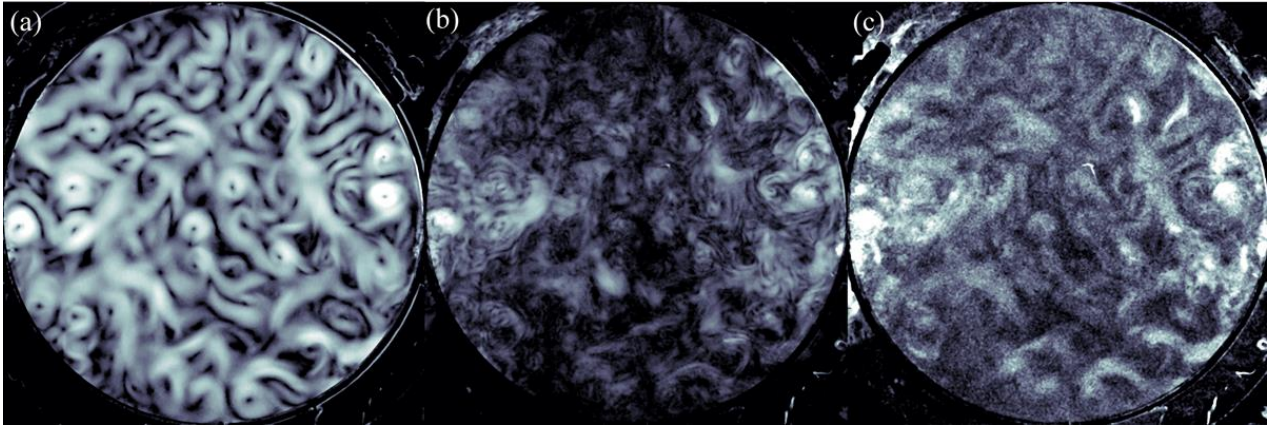

Fig. 11. Spatial distribution of energy at different frequencies in exp. 11 (Table 1): (a) zero

frequency, $\omega = 0$; (b) intermediate frequency, $\omega = 0.3f_0$; (c) high frequency, $\omega = 0.8f_0$; The greyscale shows energy in logarithmic scale.

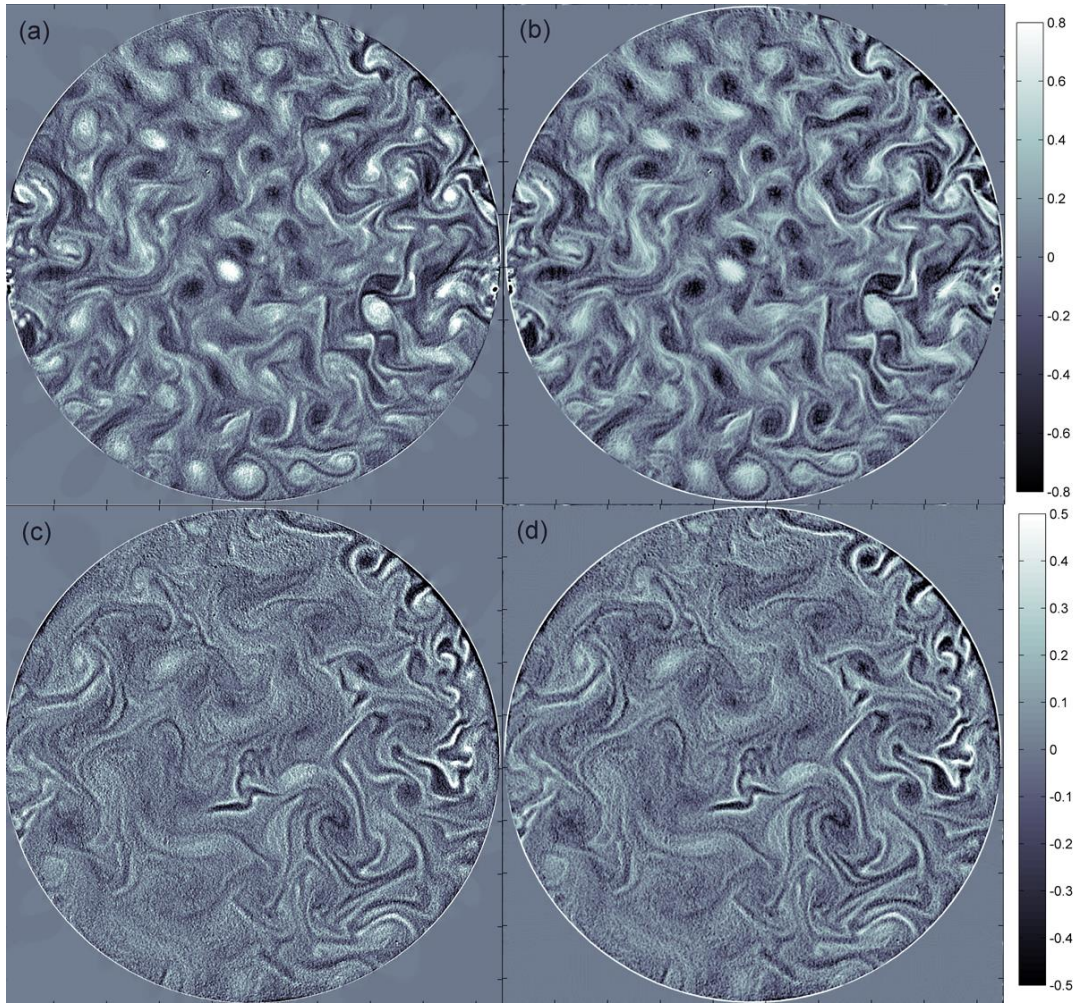

Fig. 12. Geostrophic (a, c) and total vorticity (b, d) in experiment 11 during the forcing period (a, b) and during the period of decay (c, d). Greyscale shows vorticity normalized by the Coriolis parameter $f_0$.

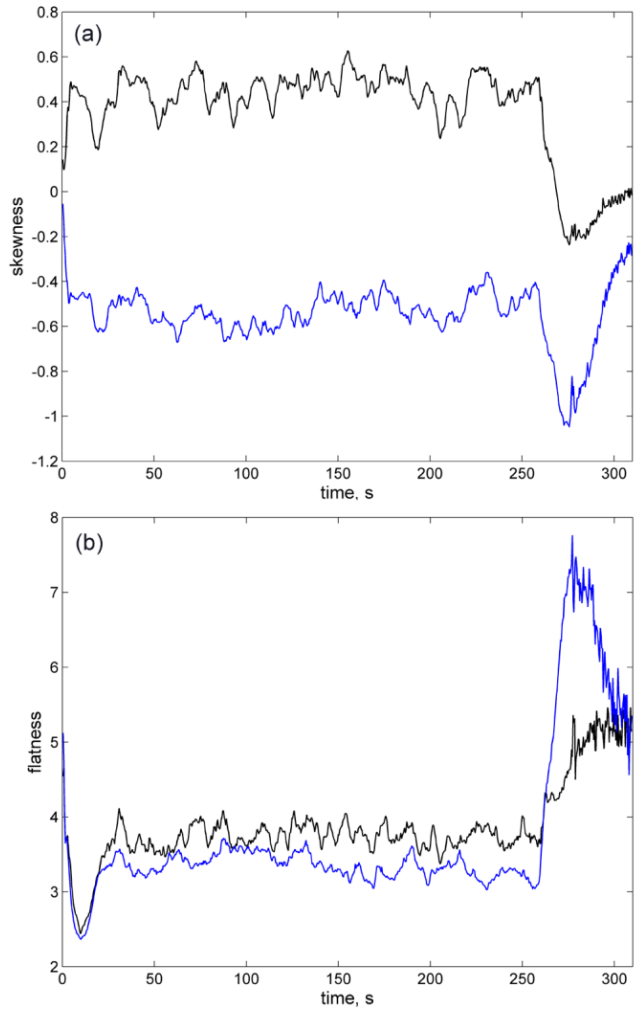

Fig. 13. (Color online) Vorticity skewness (a) and kurtosis (b) *versus* time in experiment 13 (Table 1). *S* and *K* for geostrophic and total vorticity are shown by black and blue lines respectively. Forcing is applied during the time interval $t = 0$ s to 260 s.

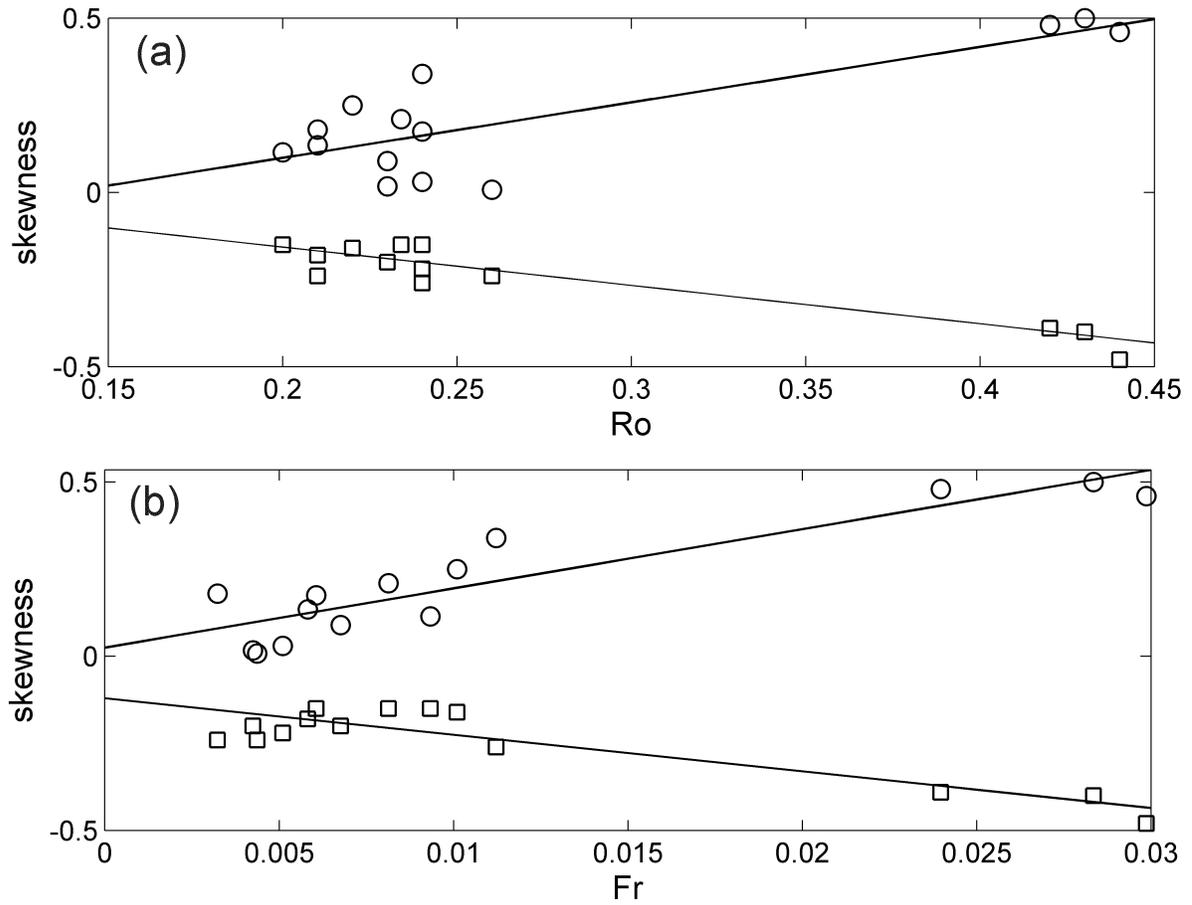

Fig. 14. Vorticity skewness as a function of Ro (a) and Fr (b) measured during the forcing period in experiments 1-14. Circles and squares show skewness calculated using geostrophic and vorticity respectively. Lines show linear fit.

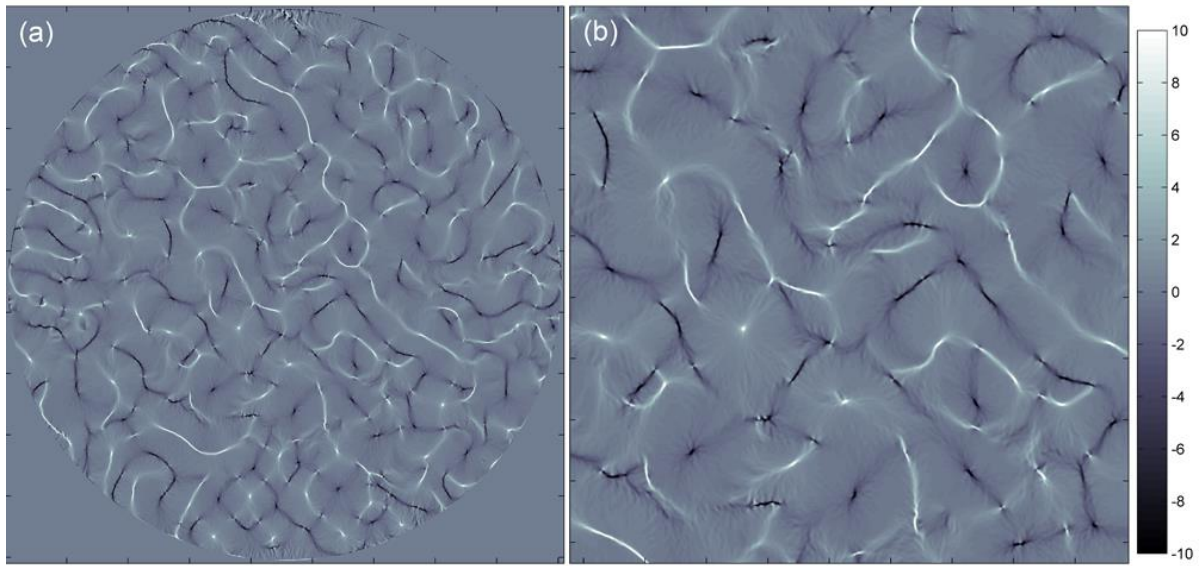

Fig. 15. Snapshot of the curvature field in experiment 13: (a) entire domain and (b) magnified area at the center. Greyscale gives κ in cm$^{-1}$ such that black and white correspond to anticyclonic and cyclonic vorticity respectively.

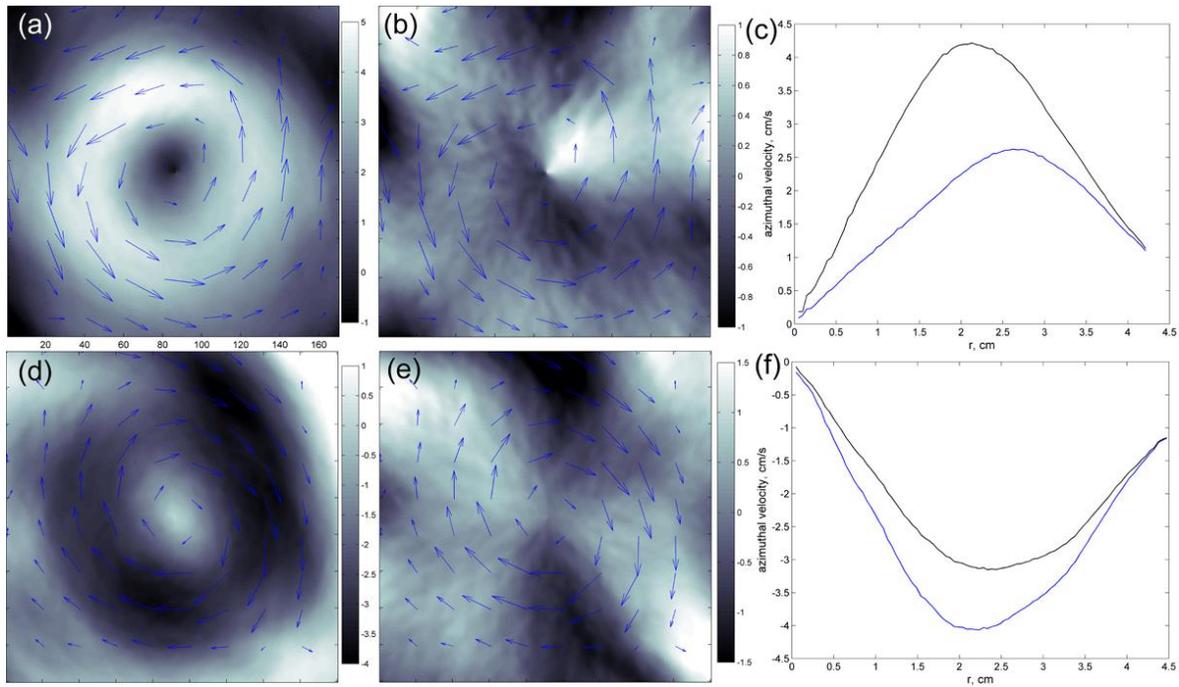

Fig. 16. (Color online) Velocity fields in a cyclone (a - c) and anticyclone (d - f) in the flow

shown in Fig.15. Greyscale gives azimuthal velocity in (a) and (d) and radial velocity in (b) and

(e). Arrows show total velocity vectors. (c) and (f) show profiles of the azimuthal velocity averaged in the azimuthal direction for each vortex. Black line shows geostrophic velocity while blue line shows the total velocity.